\documentclass[aps,prd,reprint,amsmath,amssymb,nobalancelastpage]{revtex4-2}

\usepackage{txfonts}
\DeclareMathAlphabet{\mathcal}{OMS}{cmsy}{m}{n}

\usepackage[utf8]{inputenc}
\usepackage[dvipsnames]{xcolor}
\usepackage[colorlinks=true,linkcolor=Blue,citecolor=Blue,urlcolor=Blue]{hyperref}
\usepackage{graphicx}
\usepackage[english]{isodate}
\cleanlookdateon

\newcommand{\Hess}{\operatorname{Hess}}
\newcommand{\ed}{\mathrm{d}}
\newcommand{\tr}{\operatorname{tr}}
\newcommand{\LR}{\, \big|_{\textrm{LR}}}

\begin{document}

\title{Stability conditions for light rings in circular stationary axisymmetric spacetimes}

\author{Jake O.~Shipley}
\email{shipley.jake.oliver@gmail.com}
\affiliation{Government Operational Research Service, UK Civil Service, Sheffield, United Kingdom}

\date{\today}

\begin{abstract}
	We consider the stability of light rings in four-dimensional circular stationary axisymmetric spacetimes. Using a Hamiltonian formalism for null geodesics and the Killing symmetries of the spacetime, we show that the transverse Hessian of the geodesic potential, which controls the light ring's stability, reduces there to a pure curvature expression. Its eigenvalues are $\mu_{\pm} = \Phi_{0 0} \pm \lvert \Psi_{0} \rvert$, where $\Phi_{0 0}$ is a Ricci scalar and $\Psi_{0}$ a Weyl scalar of a complex null tetrad adapted to the light ring. Through the Sachs optical equations, these quantities are responsible for focussing and shear of a null geodesic congruence. Stability requires both eigenvalues to be positive, so a light ring is stable if and only if $\Phi_{0 0} > \lvert \Psi_{0} \rvert$ there. Assuming the Einstein field equations with generic stress--energy tensor $T_{a b}$, we recast this as $4 \pi T_{a b} p^{a} p^{b} > \lvert \Psi_{0} \rvert$, where $p^{a}$ is the light ring's tangent vector: for stability, the focussing by matter must dominate the local tidal shear. The trace of the Hessian alone provides an inexpensive test of stability, requiring no knowledge of the Weyl curvature, since a light ring at which the strict null energy condition is violated cannot be stable. Applying the test to specific matter models, we recover known results for vacuum and electrovacuum, with and without a cosmological constant. We also demonstrate that stable light rings are forbidden for a minimally coupled real scalar field with any self-interaction potential, since such a field must be invariant under both isometries. We show that a complex scalar field escapes this obstruction whenever it carries harmonic dependence on the Killing coordinates.
\end{abstract}

\maketitle

\section{Introduction}
\label{sec:intro}

Since Eddington's observation of gravitational light deflection by the Sun during the total solar eclipse of May 1919 \cite{DysonEddingtonDavidson1920}, there has been considerable interest in the general-relativistic effects of lensing by compact bodies \cite{Wambsganss1998, Perlick2004}. Around black holes and other ultra-compact objects, the gravitational field can be so strong that photons are forced to follow circular orbits, known as \emph{light rings} \cite{CunhaHerdeiro2018}. Unstable light rings are intimately connected to the black hole shadow \cite{Synge1966}, as observed by the Event Horizon Telescope a century after Eddington's experiment \cite{EventHorizonTelescope2019}. In fact, horizonless ultra-compact objects may also cast a shadow, thereby mimicking the optical appearance of black holes, precisely because they admit unstable light rings \cite{CunhaEtAl2016, CunhaEtAl2017}. It has also been shown that such objects exhibit a gravitational-wave ringdown similar to that of black holes \cite{CardosoFranzinPani2016}.

\emph{Stable} light rings, which are less common than their unstable counterparts, can trap radiation in their vicinity, and are responsible for rich phenomena in spacetimes that admit them. Keir \cite{Keir2016} showed that linear waves in spacetimes exhibiting stable trapping of null geodesics cannot decay faster than logarithmically, and Cardoso \emph{et al.}~\cite{CardosoEtAl2014} argued that such slow decay is suggestive of a non-linear instability of the spacetime. Indeed, non-linear evolutions of ultra-compact Boson stars \cite{CunhaEtAl2023} confirm that the instability can destroy horizonless objects that might otherwise mimic black holes, and the backreaction of a perturbation near a stable light ring has recently been examined \cite{Cunha2025}. It is therefore of interest to know which spacetimes admit stable light rings, and a growing body of literature suggests that the answer is controlled by the matter content, i.e., the stress--energy tensor $T_{a b}$.

One approach to the study of stable light rings focusses on global topological analyses. Cunha, Berti and Herdeiro \cite{CunhaBertiHerdeiro2017} show, using the Brouwer degree of a continuous map, that the light rings of stationary axisymmetric horizonless ultra-compact objects come in pairs -- one a saddle point and the other an extremum of an effective potential. Invoking the Einstein field equations, they show that the extremum is a minimum, and hence a stable light ring, when the stress--energy tensor satisfies the null energy condition at the light ring. The argument has since been extended to black holes \cite{CunhaHerdeiro2020}, considered in the context of traversable wormholes \cite{XavierHerdeiroCrispino2024}, recast as a topological charge \cite{Wei2020}, and generalised to spacetimes with an ergoregion \cite{GhoshSarkar2021} and to non-spherical horizon topologies \cite{CunhaHerdeiroNovo2024}. Guo and Gao \cite{GuoGao2021} obtained related existence results, together with the observation that the outermost light ring of a non-extremal black hole is radially unstable and that at least one light ring is angularly stable, without invoking any energy condition. Hod \cite{Hod2018} identified an exception to the pairing theorem of Ref.~\cite{CunhaBertiHerdeiro2017} for degenerate light rings.

A complementary programme of work is local, and asks what the field equations imply about stability at a light ring. In Ref.~\cite{DolanShipley2016}, using the Einstein--Maxwell equations, it was shown that stable light rings are forbidden in four-dimensional stationary axisymmetric vacuum spacetimes but permitted in electrovacuum. In Ref.~\cite{Shipley2023}, this result was extended to include a cosmological constant. In both cases, a striking cancellation occurred: on substituting the field equations into the Hessian of the geodesic potential, the derivatives of the metric functions dropped out and only stress--energy terms remained.

Other studies have analysed \emph{photon surfaces} -- timelike hypersurfaces on which every null geodesic initially tangent remains so \cite{ClaudelVirbhadraEllis2001, Perlick2005}. Koga and Harada \cite{KogaHarada2019} analyse the stability of null geodesics on such a surface via the geodesic deviation equation, and obtain a criterion in terms of the Riemann curvature tensor contracted with the null tangent vector and the unit normal to the surface. Their framework requires no spacetime symmetry, but it does require the orbit to lie on a totally umbilic hypersurface, which the light rings of rotating spacetimes generically do not. In Kerr spacetime \cite{Teo2003}, for example, the photon spheres of constant Boyer--Lindquist radius filling the photon region are not umbilic \cite{KobialkoGaltsov2020}.

Treatments of light rings in the stationary axisymmetric case have typically been topological or geometric, making little use of the matter content, while those that exploit the field equations and energy conditions have been restricted to specific matter models \cite{DolanShipley2016, Shipley2023}, or to static spherical symmetry \cite{Hod2018}. It is natural to ask which general matter sources can, in principle, give rise to stable light rings, and which forbid them. The present study addresses this question for four-dimensional circular stationary axisymmetric spacetimes. This paper and its key results are organised as follows.

In Sec.~\ref{sec:spacetime_geom}, we summarise four-dimensional stationary axisymmetric geometries and their Killing symmetries, and introduce the concept of circularity.

Section \ref{sec:light_rings} features a review of the Hamiltonian formalism for null geodesics (Sec.~\ref{sec:hamiltonian_formalism}), and defines light rings as critical points of the norm of a helical Killing vector or, equivalently, the geodesic potential (Sec.~\ref{sec:light_ring_killing_potential}).

We study the stability of light rings in Sec.~\ref{sec:stability_conditions}. First, we introduce the transverse Hessian of the geodesic potential (Sec.~\ref{sec:transverse_hessian}), whose eigenvalues $\mu_{\pm}$ control the stability of light rings. Next, using the Hamiltonian formalism and the spacetime's Killing symmetries, we show that the transverse Hessian evaluated at a light ring reduces to a pure curvature expression (Sec.~\ref{sec:curvature_identities}). Then, introducing a complex null tetrad adapted to the light ring, we derive expressions for the Hessian's eigenvalues (Sec.~\ref{sec:scalar_invariants}). In particular, we show that
\begin{equation}
	\label{eqn:intro_eigenvalues}
	\mu_{\pm} = ( \Phi_{0 0} \pm \lvert \Psi_{0} \rvert ) \LR = ( 4 \pi \, T_{a b} \, p^{a} p^{b} \pm \lvert \Psi_{0} \rvert ) \LR ,
\end{equation}
where $\Phi_{0 0}$ and $\Psi_{0}$ are, respectively, Ricci and Weyl scalars of the Newman--Penrose formalism, and $p^{a}$ is the tangent vector to the light ring. Throughout this work, we employ the notation $\LR$ to denote that the preceding term has been evaluated at a light ring. Using these eigenvalues, we state our main result (Sec.~\ref{sec:stability_theorem}): a light ring is stable if and only if $4 \pi T_{a b} p^{a} p^{b} > \lvert \Psi_{0} \rvert$ there. The condition is algebraic and is evaluated at a single point (the light ring). It involves two scalar quantities: $\Phi_{0 0} = 4 \pi \, T_{a b} \, p^{a} p^{b}$, fixed by the matter content through the field equations, and responsible for the focussing of a null geodesic congruence; and $\Psi_{0}$, which is the pure vacuum curvature, responsible for the shear of a congruence.

In Sec.~\ref{sec:consequences}, we explore the consequences of our stability theorem. We first consider the special case of equatorial light rings (Sec.~\ref{sec:equatorial}). Then, we present a necessary criterion for stability that arises from the trace of the Hessian (Sec.~\ref{sec:trace_condition}). Finally, we apply our main result to a sequence of matter models of physical interest (Sec.~\ref{sec:matter_models}). Besides recovering known results for vacuum and electrovacuum, with and without a cosmological constant \cite{DolanShipley2016, Shipley2023}, we study scalar fields. We find that stable light rings are forbidden for a minimally coupled real scalar field with any self-interaction potential, since such a field must be invariant under both isometries and is hence constant along the light ring. A complex scalar field circumvents this restriction whenever it carries harmonic dependence on the Killing coordinates, failing only on the synchronisation condition.

Section \ref{sec:discussion} concludes with a discussion of limitations, connections to other studies, and possible extensions to this work.

We employ the following notation and conventions. The four-dimensional spacetime has metric tensor $g_{a b}$, with Lorentzian signature $(-, +, +, +)$. The inverse metric is $g^{a b}$. Lowercase Latin letters $a, b, c, \ldots$ run over spacetime indices, and the Einstein summation convention is used throughout. The covariant derivative is $\nabla_{a}$, and the partial derivative $\partial_{a}$. We use geometrised units, in which the speed of light in vacuum $c$ and the gravitational constant $G$ are set to unity.

\section{Spacetime geometry and symmetries}
\label{sec:spacetime_geom}

Consider a four-dimensional stationary axisymmetric spacetime $(\mathcal{M}, g_{a b})$ with two commuting Killing vectors $\xi^{a} \partial_{a} = \partial_{t}$ and $\eta^{a} \partial_{a} = \partial_{\phi}$, in coordinates $x^{a} \in \{ t, \phi, \rho, z \}$ adapted to the symmetries of the spacetime, where $\{ t, \phi \}$ are the coordinates in the \emph{Killing two-plane}, and $\{ \rho, z \}$ are coordinates on the \emph{transverse two-plane}. Each Killing vector satisfies \emph{Killing's equation},
\begin{equation}
	\label{eqn:killing}
	\nabla_{a} \xi_{b} + \nabla_{b} \xi_{a} = 0 , \qquad
	\nabla_{a} \eta_{b} + \nabla_{b} \eta_{a} = 0.
\end{equation}
Using these equations, coupled with the Ricci identity and the first Bianchi identity, it is straightforward to show that the Killing vectors satisfy
\begin{equation}
	\label{eqn:killing_riemann}
	\nabla_{a} \nabla_{b} \xi_{c} = R_{d a b c} \, \xi^{d} , \qquad
	\nabla_{a} \nabla_{b} \eta_{c} = R_{d a b c} \, \eta^{d} ,
\end{equation}
where $R_{a b c d}$ are the components of the Riemann curvature tensor. For each Killing vector, one may introduce the totally antisymmetric \emph{Killing bivectors}
\begin{equation}
	\label{eqn:killing_bivector_def}
	\begin{split}	
	F^{( \xi )}{}_{a b} &= 2 \nabla_{[ a} \xi_{b ]} = \nabla_{a} \xi_{b} - \nabla_{b} \xi_{a} = \partial_{a} \xi_{b} - \partial_{b} \xi_{a} , \\
	F^{( \eta )}{}_{a b} &= 2 \nabla_{[ a} \eta_{b ]} = \nabla_{a} \eta_{b} - \nabla_{b} \eta_{a} = \partial_{a} \eta_{b} - \partial_{b} \eta_{a} ,
	\end{split}
\end{equation}
where the second equality in each line follows from
Eq.~\eqref{eqn:killing} and the third from the symmetry of the torsion-free Christoffel connection. The Killing vectors may therefore be regarded as potentials for the Killing bivectors, in direct analogy with the Maxwell field; indeed, Eq.~\eqref{eqn:killing_riemann} implies that the Killing bivectors satisfy Maxwell's equations sourced by the Riemann tensor \cite{Papapetrou1966}.

The spacetime is said to be \emph{circular} if the Killing vectors satisfy \cite{Wald1984, BabichevMazza2025}
\begin{equation}
	\label{eqn:circularity_conditions}
	\xi_{[ a} \eta_{b} \nabla_{c} \xi_{d ]} = 0 , \qquad
	\eta_{[ a} \xi_{b} \nabla_{c} \eta_{d ]} = 0 .
\end{equation}
Via Killing's equation \eqref{eqn:killing} and the Einstein field equations
$G_{a b}= 8 \pi \, T_{a b}$, with generic stress--energy tensor $T_{a b}$ , this reduces to the requirement that
\begin{equation}
	\label{eqn:circularity_components}
	T_{t\rho} = T_{t z} = T_{\phi \rho} = T_{\phi z} = 0 ,
\end{equation}
together with the other vanishing components obtained by symmetry ($T_{a b} = T_{b a}$).

Circularity is automatic for the paradigmatic Kerr solution \cite{Papapetrou1966}, and indeed for any stationary axisymmetric \emph{electrovacuum} solution in general relativity, for which the Maxwell field shares the spacetime symmetries \cite{KundtTrumper1966, Wald1984, Heusler1996, ChruscielLopesCostaHeusler2012}. However, more exotic matter models, such as flows or fields in the meridional plane \cite{GourgoulhonBonazzola1993}, are non-circular while retaining stationarity and axial symmetry. In this work we impose circularity as an additional hypothesis.

For circular spacetimes, the line element can be expressed in block-diagonal (``orthogonally transitive'') form as \cite{Carter1969, Wald1984}
\begin{equation}
	\begin{split}
		\label{eqn:sa_metric}
		\ed s^{2} &= g_{a b} \, \ed x^{a} \ed x^{b} \\
		&= - U \, \ed t^{2} + 2 W  \, \ed t  \, \ed \phi + X \, \ed \phi^{2} + e^{2 \lambda} ( \ed \rho^{2} + \ed z^{2} ) ,
	\end{split}
\end{equation}
where $U$, $W$, $X$ and $\lambda$ are all functions of only the isothermal coordinates $\rho$ and $z$. The only non-zero components of the inverse metric tensor $g^{a b}$ are $g^{t t} = - X / B^{2}$, $g^{t \phi} = W / B^{2}$, $g^{\phi \phi} = U / B^{2}$, $g^{\rho \rho} = g^{z z} = e^{- 2 \lambda}$, where
\begin{equation}
	\label{eqn:B_defn}
	B^2 = U X + W^2 ,
\end{equation}
and $-B^{2}$ is the determinant of the $(t, \phi)$-block of $g_{a b}$. We take $B > 0$ and note that $B$ vanishes on the rotation axis. The metric functions in the $(t, \phi)$-block can be expressed as inner products of the Killing vectors: $U = - g_{t t} = - \xi^{a} \xi_{a}$, $X = g_{\phi \phi} = \eta^{a} \eta_{a}$ and $W = g_{t \phi} = \xi^{a} \eta_{a}$.

The projection operator onto the Killing two-plane is
\begin{equation}
	\label{eqn:killing_projector}
	Q_{ab} = \frac{1}{B^{2}}
	( -X \, \xi_{a} \xi_{b} + 2 W \, \xi_{ ( a } \eta_{b ) } + U \, \eta_{a} \eta_{b} ),
\end{equation}
and the complementary projector onto the orthogonal transverse two-plane is
\begin{equation}
	\label{eqn:transverse_projector}
	P_{a b } = g_{ab} - Q_{a b} .
\end{equation}
The projection operator \eqref{eqn:killing_projector} satisfies
$Q_{a b} \xi^{b} = \xi_{a}$, $Q_{a b} \eta^{b} = \eta_{a}$ and $Q_{a c} Q^{c}{}_{b} = Q_{a b}$, so the transverse projector \eqref{eqn:transverse_projector} satisfies the properties
\begin{equation}
	\label{eqn:transverse_projector_properties}
	P_{a c} P^{c}{}_{b} = P_{a b} , \qquad
	P^{a}{}_{a} = 2, \qquad
	P_{a b} \xi^{b} = 0 = P_{a b} \eta^{b} .
\end{equation}
In adapted coordinates $\{ t, \phi, \rho, z \}$, $P_{a b}$ has components $P_{\rho \rho} = e^{2 \lambda} = P_{z z}$ and all others zero.

\section{Light rings}
\label{sec:light_rings}

\subsection{Hamiltonian formalism for null geodesics}
\label{sec:hamiltonian_formalism}

Geodesics of the spacetime $\eqref{eqn:sa_metric}$ are integral curves $x^{a}(s)$ of Hamilton's equations with Hamiltonian $H = \frac{1}{2} g^{a b} p_{a} p_{b}$, where $p_{a} = g_{a b} p^{b} = g_{a b} \dot{x}^{b}$ are the canonical momenta, with an overdot denoting differentiation with respect to the geodesic's affine parameter $s$. Due to the existence of the spacetime's Killing objects (including the metric, which is itself a trivial rank-two Killing tensor), the quantities $H$, $p_{t} = \xi^{a} p_{a}$ and $p_{\phi} = \eta^{a} p_{a}$ are all conserved along geodesics, with $H = 0$ in the null case. Further, one may set $p_{t} = -1$ without loss of generality by using the freedom to rescale the affine parameter along null geodesics. Explicitly, the Hamiltonian reads
\begin{equation}
	\label{eqn:hamiltonian_explicit}
	H = \frac{1}{2} ( P^{a b} + Q^{a b} ) p_{a} p_{b} = \frac{1}{2} e^{- 2 \lambda} ( p_{\rho}^{2} + p_{z}^{2} ) + V ,
\end{equation}
with null geodesic potential
\begin{equation}
	\label{eqn:potential_explicit}
	V = \frac{1}{2} Q^{a b} p_{a} p_{b} = - \frac{1}{2 B^{2}} ( X  + 2 W p_{\phi} - U p_{\phi}^{2} ) .
\end{equation}

Since $B^{2} = U X + W^{2}$, the potential \eqref{eqn:potential_explicit} may be factorised as
\begin{equation}
	\label{eqn:quadratic_factorized}
	V = - \frac{U}{2 B^{2}} ( h^{+} - p_{\phi} ) ( h^{-} + p_{\phi} ) ,
\end{equation}
via the introduction of the effective potentials
\begin{equation}
	\label{eqn:effective_potentials}
	h^{\pm} = \frac{B \pm W}{U} ,
\end{equation}
following the approach laid out in Refs.~\cite{DolanShipley2016, Shipley2019, Shipley2023} and employed elsewhere. In those analyses, it was convenient to work with the effective potentials \eqref{eqn:effective_potentials}, which are independent of the parameter $p_{\phi}$. However, in this analysis, we favour the geodesic potential \eqref{eqn:potential_explicit}.

\subsection{Light rings as stationary points of a Killing norm and the geodesic potential}
\label{sec:light_ring_killing_potential}

Light rings are circular null geodesics of constant $\rho$ and $z$, with $p_{\rho} = 0 = p_{z}$ (as well as $p_{t} = -1$ and $p_{\phi}$ constant, as for all null rays). A light ring therefore has constant angular velocity $\Omega = \ed \phi / \ed t \neq 0$, and its tangent vector $p^{a} = g^{a b} p_{b}$ is necessarily proportional to the helical Killing vector
\begin{equation}
	\label{eqn:chi_def}
	\chi^{a} = \xi^{a} + \Omega \, \eta^{a} ,
\end{equation}
which is manifestly future-pointing. (This is a linear combination of the two Killing vectors $\xi^{a}$ and $\eta^{a}$, which is itself a Killing vector.)

We note three important properties of the composite Killing vector \eqref{eqn:chi_def}. First, since the light ring is null, $\chi^{a} \chi_{a} \LR = 0$ and hence $\chi^{a} p_{a} \LR = 0$. We have $\chi^{a} p_{a} = p_{t} + \Omega \, p_{\phi} = -1 + \Omega \, p_{\phi}$ from the conserved quantities, so $\Omega = 1 / p_{\phi}$ at a light ring. Second, the norm of $\chi^{a}$ is
\begin{equation}
	\label{eqn:norm_chi}
	\chi^{a} \chi_{a} = - U + 2 \Omega W + \Omega^{2} X = - 2 \Omega^{2} B^{2} V ,
\end{equation}
where $V$ is the geodesic potential \eqref{eqn:potential_explicit}. Since $\Omega^{2} B^{2} > 0$, the geodesic potential \eqref{eqn:potential_explicit} is, up to a negative factor, equivalent to the norm of the Killing vector \eqref{eqn:chi_def}. At the light ring, $\chi^{a}$ is null, so its norm vanishes; this is equivalent to the condition $V = 0$ by Eq.~\eqref{eqn:norm_chi}. Third, since $\chi^{a}$ is a Killing vector, $\chi^{a} \nabla_{a} \chi_{b} = -\tfrac{1}{2} \nabla_{b} (\chi^{a} \chi_{a})$. The fact that the ray is an affinely parameterised geodesic makes the left-hand side vanish at a light ring (via the geodesic equation), which means that the right-hand side must also vanish. This means $\nabla_{a} (\chi^{b} \chi_{b}) = 0$.

A light ring is therefore characterised by the covariant conditions
\begin{equation}
	\label{eqn:covariant_light_ring_conditions}
	\chi^{a} \chi_{a} = 0 , \qquad \nabla_{a} (\chi^{b} \chi_{b}) = 0 .
\end{equation}
In other words, light rings are critical points of the norm of the Killing vector $\chi^{a}$, which is proportional to the tangent vector at the light ring.

In more familiar language, using the geodesic potential \eqref{eqn:potential_explicit} that appears naturally in the Hamiltonian formalism, these conditions are equivalent to
\begin{equation}
	\label{eqn:potential_light_ring_conditions}
	V = 0 , \qquad \nabla_{a} V = \partial_{a} V = 0 ,
\end{equation}
where the covariant derivative can be replaced by a partial derivative in the second condition by virtue of the fact that $V$ is a scalar function. The first of these conditions can be seen from the Hamiltonian constraint $H = 0$ with $p_{\rho} = 0 = p_{z}$. The conditions \eqref{eqn:potential_light_ring_conditions} say that light rings are simultaneously zeros and stationary points of the potential \eqref{eqn:potential_explicit}. As shown in Ref.~\cite{DolanShipley2016} and elsewhere, light rings are also stationary points of the effective potentials \eqref{eqn:effective_potentials}, with $h^{\pm} = \pm p_{\phi}$ and $\partial_{a} h^{\pm} = 0$ there.

As $p^{a}$ is proportional to $\chi^{a}$ at the light ring, we may write $p^{a} = N \chi^{a}$ for some normalisation factor $N$, to be determined. Contracting with $\xi^{a}$ gives $-1 = - N ( U - \Omega W )$. Squaring both sides and using the null condition (which permits us to make the replacement $U - 2 \Omega W = \Omega^{2} X $), and the definition in Eq.~\eqref{eqn:B_defn}, we find $1 = N^{2} \Omega^{2} ( U X + W^{2} ) = N^{2} \Omega ^{2} B^{2}$. Hence $N = \varepsilon/ (\Omega B)$, with $\varepsilon = \operatorname{sign} ( \Omega) = \pm 1$. Therefore,
\begin{equation}
	\label{eqn:tangent_killing}
	p^{a} = \frac{\varepsilon}{\Omega B}  ( \xi^{a} + \Omega \, \eta^{a} ) = \frac{\varepsilon}{B} (p_{\phi} \, \xi^{a} + \eta^{a} ) ,
\end{equation}
with the two signs corresponding to prograde ($\varepsilon = 1$) and retrograde ($\varepsilon = -1$) branches, $p_{\phi} = h^{+}$ and $p_{\phi} = - h^{-}$, respectively. One can see from the definition of $\varepsilon$ and the coefficient of $\xi^{a}$ that the tangent vector \eqref{eqn:tangent_killing} is future-pointing.

It is clear from the fact that the tangent vector at the light ring \eqref{eqn:tangent_killing} is a linear combination of Killing vectors $\xi^{a}$ and $\eta^{a}$ that it lies in the Killing two-plane. (This is not true for a general null geodesic, which has at least one of $p_{\rho}$ and $p_{z}$ non-zero.) This fact can also be seen via the use of the projection operators \eqref{eqn:killing_projector} and \eqref{eqn:transverse_projector}: $Q_{a b} p^{b} = p_{a}$ and $P_{a b} p^{b} = 0$ at the light ring.

\section{Stability conditions for light rings}
\label{sec:stability_conditions}

\subsection{Transverse Hessian and stability criteria}
\label{sec:transverse_hessian}

Perturbing a null geodesic from a light ring in the transverse plane by $\delta x^{a}$, Hamilton's equations for Eq.~\eqref{eqn:hamiltonian_explicit} give $\dot{x}^{a} = P^{a b} p_{b}$ and $\dot{p}_{a} = -\nabla_{a} V$. The transverse displacement therefore obeys
\begin{equation}
	\label{eqn:linear_perturbation}
	\delta \ddot{x}^{a} = -\mathcal{H}^{a}{}_{b} \LR \, \delta x^{b} ,
\end{equation}
where we have introduced the \emph{transverse Hessian} of the geodesic potential $V$,
\begin{equation}
	\label{eqn:transverse_hessian}
	\mathcal{H}_{ab} = P_{a}{}^{c} \, P_{b}{}^{d} \, \nabla_{c} \nabla_{d} V .
\end{equation}
This the standard covariant Hessian of the full four-dimensional spacetime $[\Hess V]_{a b} = \nabla_{a} \nabla_{b} V$, projected onto the transverse two-plane. The transverse Hessian \eqref{eqn:transverse_hessian} is a symmetric spacetime tensor, defined without reference to any coordinate system or index restriction. It annihilates the Killing directions: $\mathcal{H}_{ab} \xi^{b} = 0 = \mathcal{H}_{ab} \eta^{b}$. As an operator, $\mathcal{H}^{a}{}_{b}$ has eigenvalues $\{ 0, 0, \mu_{+}, \mu_{-} \}$, where the two non-trivial eigenvalues are equal to those of the intrinsic Hessian on the transverse plane.

The projection operator $P^{a b}$ is positive definite (recall that its only non-zero components are $P^{\rho \rho} = e^{- 2 \lambda} = P^{z z}$). It follows from Eq.~\eqref{eqn:linear_perturbation} that the light ring is stable if and only if both non-trivial eigenvalues of $\mathcal{H}^{a}{}_{b}$ are strictly positive: $\mu_{\pm} > 0$. Let us first introduce the scalar invariants
\begin{align}
	\label{eqn:trace}
	\tr{\mathcal{H}} &= g^{a b} \mathcal{H}_{a b} = \mathcal{H}^{a}{}_{a} = \mu_{+} + \mu_{-} , \\
	\label{eqn:determinant}
	\det{\mathcal{H}} &= \tfrac{1}{2} \left[ ( \tr \mathcal{H} )^{2} - \mathcal{H}_{a b} \mathcal{H}^{a b} \right] = \mu_{+} \mu_{-} ,
\end{align}
where the relationship between the determinant and trace in Eq.~\eqref{eqn:determinant} is derived using the Cayley--Hamilton theorem for $\mathcal{H}_{ab}$, which is a degenerate rank-two tensor on the full four-dimensional spacetime manifold, and its restriction to the image of the projector $P_{ab}$ defines a non-degenerate two-dimensional tensor on the transverse plane. It is therefore understood that Eq.~\eqref{eqn:determinant} represents the intrinsic two-dimensional determinant on the transverse plane. Using these invariants, the light ring stability criteria are simply
\begin{align}
	\label{eqn:stable_trace}
	\tr{\mathcal{H}} \LR &> 0 , \\
	\label{eqn:stable_determinant}
	\det{\mathcal{H}} \LR &> 0 .
\end{align}
If $\det{\mathcal{H}} \LR > 0$ but $\tr{\mathcal{H}} \LR < 0$, the the critical point is a local maximum of $V$, hence the light ring is unstable. If $\det{\mathcal{H}} \LR < 0$, the light ring is a saddle and hence unstable. If $\det{\mathcal{H}} \LR = 0$, it is marginal and higher-order terms must be retained; this is beyond the scope of the present work.

\subsection{Curvature identity for the transverse Hessian at a light ring}
\label{sec:curvature_identities}

In the remainder of this section, we will use the spacetime's symmetry and curvature properties to derive generic expressions for the components of the transverse Hessian \eqref{eqn:transverse_hessian}. This will allow us to recast the stability conditions \eqref{eqn:stable_trace} and \eqref{eqn:stable_determinant} as algebraic identities at the light ring.

As the potential \eqref{eqn:potential_explicit} is a scalar function, it satisfies $\nabla_{a} V = \partial_{a} V$ and $\nabla_{a} \nabla_{b} V = \partial_{a} \partial_{b} V - \Gamma^{c}{}_{a b} \, \partial_{c} V$. Since $\partial_{a} V = 0$ at the light ring, we have
\begin{equation}
	\label{eqn:partial_covariant}
	\partial_{a} V \LR = \nabla_{a} V \LR = 0 , \qquad
	\partial_{a} \partial_{b} V \LR = \nabla_{a} \nabla_{b} V \LR .
\end{equation}
These identities mean that covariant derivatives and partial derivatives of $V$ are interchangeable at the light ring. As we shall see, working in terms of covariant derivatives will permit us to express the components of the Hessian in terms of spacetime curvature invariants, which arise generically from the definition of $\nabla_{a}$.

Employing the null condition at the light ring, we find that the covariant derivative of the geodesic potential satisfies
\begin{equation}
	\label{eqn:covariant_potential_light_ring}
	0 = \nabla_{a} V \LR = - \frac{1}{2 B^{2}} \nabla_{a} ( X  + 2 W p_{\phi} - U p_{\phi}^{2} ) \LR .
\end{equation}
Similarly, we find that the second covariant derivative of $V$ evaluated at a light ring is
\begin{gather}
	\label{eqn:covariant_potential}
	\begin{aligned}
	\nabla_{a} \nabla_{b} V \LR &= - \frac{1}{2 B^{2}} \nabla_{a} \nabla_{b} ( X  + 2 W p_{\phi} - U p_{\phi}^{2} ) \LR \\
	&= - \frac{1}{2 B^{2}} ( \nabla_{a} \nabla_{b} X  + 2 p_{\phi} \, \nabla_{a} \nabla_{b} W - p_{\phi}^{2} \, \nabla_{a} \nabla_{b} U ) \LR
	.
	\raisetag{41pt}
	\end{aligned}
\end{gather}
where $p_{\phi}$ is constant along null geodesics, so it can be passed through the covariant derivatives.

We now seek expressions for the terms on the right-hand side of Eq.~\eqref{eqn:covariant_potential}, which are second covariant derivatives of metric functions in the $(t, \phi)$-block. For this purpose, let us return to the symmetry properties of the spacetime reviewed in Sec.~\ref{sec:spacetime_geom}. In particular, we recall that the metric components of interest can be written as inner products of Killing vectors. Applying the covariant derivative twice to the functions $U = - \xi^{a} \xi_{a}$, $X = \eta^{a} \eta_{a}$ and $W = \xi^{a} \eta_{a}$, which are all bilinear in the Killing vectors, and using Eqs.~\eqref{eqn:killing_riemann} and \eqref{eqn:killing_bivector_def}, yields the identities
\begin{align}
	\label{eqn:U_bilinear}
	\nabla_{a} \nabla_{b} U &= - \tfrac{1}{2} F^{(\xi)}{}_{b c} F^{(\xi)}{}_{a}{}^{c} - 2 R_{d a b c} \, \xi^{c} \xi^{d} , \\
	\label{eqn:X_bilinear}
	\nabla_{a} \nabla_{b} X &= \phantom{-} \tfrac{1}{2} F^{(\eta)}{}_{b c} F^{(\eta)}{}_{a}{}^{c} + 2 R_{d a b c} \, \eta^{c} \eta^{d} , \\
	\begin{split}
	\label{eqn:W_bilinear}
	\nabla_{a} \nabla_{b} W &= \phantom{-} \tfrac{1}{4} ( F^{(\xi)}{}_{b c} F^{(\eta)}{}_{a}{}^{c} + F^{(\eta)}{}_{b c} F^{(\xi)}{}_{a}{}^{c} ) \\
	& \qquad + R_{d a b c} ( \xi^{c} \eta^{d} + \eta^{c} \xi^{d} ) .
	\end{split}
\end{align}
Substituting these into Eq.~\eqref{eqn:covariant_potential}, the curvature (Riemann) and ``field-strength'' (Killing bivector) terms can be collected together to give
\begin{equation}
	\label{eqn:covariant_potential_2}
	\nabla_{a} \nabla_{b} V \LR = \left. \left( R_{a c b d} \, p^{c} p^{d} - \frac{1}{4 \Omega^{2} B^{2}} F_{b c} F_{a}{}^{c} \right) \, \right|_{\textrm{LR}} .
\end{equation}

In the second term on the right-hand side, we recall that $\Omega = 1 / p_{\phi}$, and we have introduced the quantity $F_{a b} = F^{(\xi)}{}_{a b} + \Omega \, F^{(\eta)}{}_{a b}  = 2 \nabla_{ [ a } \chi_{ b ] }$, which is the Killing bivector corresponding to the helical Killing vector \eqref{eqn:chi_def}. We recall from Sec.~\ref{sec:light_ring_killing_potential} that, at the light ring, $\chi^{a}$ is proportional to the tangent vector \eqref{eqn:tangent_killing}.

In the Riemann terms, we have replaced the Killing vectors with the light ring's tangent vector using the relation \eqref{eqn:tangent_killing}, noting that $\varepsilon$ disappears from the result because the quantity is quadratic in the tangent vector $p^{a}$, and $\varepsilon^{2} = 1$. That is, the expression is independent of the orientation of the light ring. In addition, we have changed the sign of this term using the symmetry properties of the Riemann tensor, interchanging indices via $R_{d a b c} = - R_{a d b c}$, and relabelling indices ($c \leftrightarrow d$) in the contraction with $p^{c} p^{d}$.

We now show that
\begin{equation}
	\label{eqn:vanishing_field_strength}
	P_{a}{}^{c} P_{b}{}^{d} F_{c e} F_{d}{}^{e} \LR = 0 ,
\end{equation}
using Killing's equation, the light ring conditions, and circularity. One may also derive this result in a more direct fashion using the components of the bivectors \eqref{eqn:killing_bivector_def} in coordinate form, which can be expressed as combinations of partial derivatives of the metric functions $U$, $X$ and $W$.

Firstly, Killing's equation gives $F_{a b} = 2 \nabla_{a} \chi_{b}$, so
\begin{equation}
	F_{ab} \, \chi^{b} = 2 \chi^{b} \nabla_{a} \chi_{b} = \nabla_{a} ( \chi^{b} \chi_{b} ) .
\end{equation}
The light ring condition $\nabla_{a} ( \chi^{b} \chi_{b}) \LR = 0$ therefore means that $\chi^{a}$, and hence $p^{a} \propto \chi^{a}$, are annihilated by $F_{ab}$ at the light ring
\begin{equation}
	\label{eqn:kernel}
	F_{a b} p^{b} \LR = 0 .
\end{equation}

Returning to the left-hand side of Eq.~\eqref{eqn:vanishing_field_strength}, we can insert the inverse metric to lower all indices of the Killing bivector, then use the completeness relation for the inverse metric $g^{a b} = P^{a b} + Q^{a b}$ to decompose the contraction into transverse and Killing components:
\begin{equation}
	\label{eqn:killing_transverse_split}
	P_{a}{}^{c} P_{b}{}^{d} F_{c e} F_{d}{}^{e} = P_{a}{}^{c} P_{b}{}^{d} F_{c e} F_{d f} ( P^{e f} + Q^{e f} ) .
\end{equation}

After expanding brackets, the first term on the right-hand side of Eq.~\eqref{eqn:killing_transverse_split}, which features $P^{e f}$, isolates the purely transverse components of the bivector $F_{a b} = 2\nabla_{ [ a }\chi_{ b ] }$. Let $u^{a}$ and $\varv^{a}$ be any two vector fields tangent to the transverse plane. By Eq.~\eqref{eqn:transverse_projector_properties}, the Killing vectors are orthogonal to this plane, so $\chi_{a} u^{a} = 0 = \chi_{a} \varv^{a}$. Differentiating the latter along $u^{a}$ using the product rule gives $u^{a} \varv^{b} \nabla_{a} \chi_{b} = - \chi_{b} \, u^{a} \nabla_{a} \varv^{b}$. Subtracting the same identity with $u^{a}$ and $\varv^{a}$ interchanged yields
\begin{equation}
	\label{eqn:bracket}
	F_{a b} \, u^{a} \varv^{b} = -\chi_{a} \, [u , \varv]^{a} ,
\end{equation}
where $[u , \varv]^{a}$ is the Lie bracket of $u^{a}$ and $\varv^{a}$. The transverse--transverse block of the Killing bivector -- the left-hand side of Eq.~\eqref{eqn:bracket} -- is therefore precisely the obstruction to the transverse two-planes closing under the Lie bracket. Circularity is the statement that field of transverse two-planes is integrable, so that by the Frobenius integrability theorem $[u, \varv]^{a}$ is also transverse \cite{Wald1984, Heusler1996}; hence its contraction with $\chi_{a}$ is zero, causing the right-hand side of Eq.~\eqref{eqn:bracket} to vanish. As $u^{a}$ and $\varv^{a}$ were arbitrary,
\begin{equation}
	P_{a}{}^{c} P_b{}^{d} F_{c d} = 0 ,
\end{equation}
which implies $P_{a}{}^{c} P_{b}{}^{d} F_{c e} F_{d f} P^{e f} = 0$. Applying Eq.~\eqref{eqn:bracket} to $\xi^{a}$ and $\eta^{a}$ separately shows why both circularity conditions in Eq.~\eqref{eqn:circularity_conditions} are needed: each removes one Killing-plane component of $[u, \varv]^{a}$.

For the second term of Eq.~\eqref{eqn:killing_transverse_split}, which features $Q^{e f}$, let us express the Killing projector in a null basis adapted to the light ring. The tangent vector $p^{a}$ is null and satisfies $p_{a} \xi^{a} = p_{t} = -1$ by the normalisation of the affine parameter fixed in Sec.~\ref{sec:hamiltonian_formalism}, so the Killing two-plane is non-degenerate at the light ring. The choice
\begin{equation}
	\label{eqn:n_def}
	n^{a} = \xi^{a} - \frac{U}{2} p^{a} , \qquad
	n^{a} n_{a} = 0 , \qquad 
	p^{a} n_{a} = -1 ,
\end{equation}
where $U = -\xi^{a} \xi_{a}$, completes $\{ p^{a}, n^{a} \}$ to a null basis for the Killing two-plane. In terms of this null basis, the Killing projector is
\begin{equation}
	\label{eqn:Q_decomposition}
	Q^{a b} = - p^{a} n^{b} - n^{a} p^{b} .
\end{equation}
Using this form of the projection operator, we find
\begin{equation}
	\begin{split}
	P_{a}{}^{c} P_{b}{}^{d} F_{c e} F_{d f} Q^{e f} \LR = & - ( P_{a}{}^{c} P_{b}{}^{d} F_{c e} F_{d f} p^{e} n^{f} \\
	&\quad + P_{a}{}^{c} P_{b}{}^{d} F_{c e} F_{d f} n^{e} p^{f} ) \LR .
	\end{split}
\end{equation}
Since every term on the right-hand side vanishes at the light ring by the condition \eqref{eqn:kernel}, Eq.~\eqref{eqn:vanishing_field_strength} holds.

Applying $P_{a}{}^{c} P_{b}{}^{d}$ to Eq.~\eqref{eqn:covariant_potential_2} therefore eliminates the field-strength terms entirely, leaving
\begin{equation}
	\label{eqn:hessian_result}
	\mathcal{H}_{a b} \LR = P_{a}{}^{c} P_{b}{}^{d} \, R_{c e d f} \, p^{e} p^{f} \LR .
\end{equation}

\subsection{Scalar invariants}
\label{sec:scalar_invariants}

Our expression for the components of the transverse Hessian at a light ring \eqref{eqn:hessian_result} permits us to reduce the stability criteria to pure algebra, without any derivatives of spacetime functions. In this section, we will demonstrate that both scalar invariants \eqref{eqn:trace} and \eqref{eqn:determinant} follow from a single expansion of the Hessian.

First, we construct a Newman--Penrose complex null tetrad along the light ring. Let $m^{a}$ be a complex null vector spanning the transverse plane, normalised by
\begin{equation}
	\label{eqn:m_normalisation}
	m^{a} m_{a} = 0 , \qquad
	m^{a} \bar{m}_{a} = 1 , \qquad
	P_{a b} = m_{a} \bar{m}_{b} + \bar{m}_{a} m_{b} ,
\end{equation}
so that $m^{a} p_{a} = 0$ automatically, since the tangent vector $p^{a}$ lies in the Killing two-plane. In adapted coordinates one may take $m^{a} \partial_{a} = ( e^{-\lambda} / \sqrt{2} ) ( \partial_{\rho} + i \, \partial_{z} )$. Together with $p^{a}$ and the auxiliary null vector $n^{a}$ introduced in Eq.~\eqref{eqn:n_def}, the set $\{ p^{a} , n^{a} , m^{a} , \bar{m}^{a} \}$ constitutes a complex null tetrad at the light ring, in which the completeness relation for the inverse metric reads
\begin{equation}
	\label{eqn:tetrad_completeness}
	g^{a b} = Q^{a b} + P^{a b} = - ( p^{a} n^{b} + n^{a} p^{b} ) + ( m^{a} \bar{m}^{b} + \bar{m}^{a} m^{b} ) .
\end{equation}

The transverse Hessian is a real symmetric tensor supported entirely on the transverse plane, so it can be written
\begin{equation}
	\label{eqn:hessian_expansion}
	\mathcal{H}_{a b} = \Phi \, ( m_{a} \bar{m}_{b} + \bar{m}_{a} m_{b} ) + \Psi \, \bar{m}_{a} \bar{m}_{b} + \bar{\Psi} \, m_{a} m_{b} ,
\end{equation}
where the two coefficients are recovered by contraction,
\begin{equation}
	\label{eqn:hessian_coefficients}
	\Phi = \mathcal{H}_{a b} \, m^{a} \bar{m}^{b} , \qquad \Psi = \mathcal{H}_{a b} \, m^{a} m^{b} ,
\end{equation}
using the normalisation conditions \eqref{eqn:m_normalisation} to isolate each term. Clearly $\Phi = \bar{\Phi}$, so $\Phi$ is real, whilst $\Psi$ is complex in general; the expansion \eqref{eqn:hessian_expansion} therefore carries three real parameters, as required for a real symmetric tensor on a two-dimensional plane. The sum of the final two terms in the expansion \eqref{eqn:hessian_expansion} is manifestly real, each term being the conjugate of the other.

Acting on a transverse tensor, the inverse metric may be replaced by $P^{a b}$, so the trace \eqref{eqn:trace} is simply
\begin{equation}
	\label{eqn:trace_phi}
	\tr{\mathcal{H}} = 2 \Phi .
\end{equation}
Squaring Eq.~\eqref{eqn:hessian_expansion}, every contraction reduces to $m^{a} m_{a} = 0$ or $m^{a} \bar{m}_{a} = 1$ by the normalisation \eqref{eqn:m_normalisation}; the cross terms between $\Phi$ and $\Psi$ vanish, and we are left with
\begin{equation}
	\label{eqn:H_squared}
	\mathcal{H}_{a b} \mathcal{H}^{a b} = 2 ( \Phi^{2} + \lvert \Psi \rvert^{2} ) .
\end{equation}
Substituting Eqs.~\eqref{eqn:trace_phi} and \eqref{eqn:H_squared} into the Cayley--Hamilton relation \eqref{eqn:determinant} gives the determinant of the transverse Hessian in terms of the same two scalars:
\begin{equation}
	\label{eqn:determinant_phi_psi}
	\det{\mathcal{H}} = \Phi^{2} - \lvert \Psi \rvert^{2} .
\end{equation}
Both stability criteria are therefore controlled by the pair \eqref{eqn:hessian_coefficients}.

\subsubsection{Trace}

Let us compute the trace of the transverse Hessian directly; this yields one of the scalar invariants via Eq.~\eqref{eqn:trace_phi}. From Eqs.~\eqref{eqn:trace} and \eqref{eqn:hessian_result},
\begin{equation}
	\label{eqn:trace_expansion}
	\begin{split}
		\tr{\mathcal{H}} \LR &= g^{a b} P_{a}{}^{c} P_{b}{}^{d} \, R_{c e d f} \, p^{e} p^{f} \LR \\
		&= P_{a}{}^{c} P^{a d} \, R_{c e d f} \, p^{e} p^{f} \LR \\
		&= P^{c d} R_{c e d f} \, p^{e} p^{f} \LR \\
		&= ( g^{c d} - Q^{c d} ) R_{c e d f} \, p^{e} p^{f} \LR ,
	\end{split}
\end{equation}
where we have used one of the properties of the transverse projector \eqref{eqn:transverse_projector_properties} in the third equality, and the replacement \eqref{eqn:transverse_projector} in the fourth. Expanding the right-hand side, the first term gives $g^{c d} R_{c e d f} = R_{e f}$ (the Ricci tensor) by the standard contraction. In the second term, we can replace $Q^{c d}$ using the null-basis form \eqref{eqn:Q_decomposition}. Consequently every term of $Q^{c d} R_{c e d f} p^{e} p^{f}$ carries $p^{a}$ in both slots of one antisymmetric index pair of the Riemann tensor, so the term vanishes identically. Hence
\begin{equation}
	\label{eqn:trace_ricci}
	\tr{\mathcal{H}} \LR = R_{a b} \, p^{a} p^{b} \LR , \qquad \Phi \LR = \Phi_{0 0} \LR ,
\end{equation}
where $\Phi_{0 0} = \tfrac{1}{2} R_{a b} p^{a} p^{b}$ is one of the Ricci scalars of the Newman--Penrose formalism, defined with respect to the complex null tetrad $\{ p^{a} , n^{a} , m^{a} , \bar{m}^{a} \}$.

Introducing the trace-reversed field equations $R_{a b} = 8 \pi ( T_{a b} - \tfrac{1}{2} g_{a b} T )$ and using the nullity of $p^{a}$ to eliminate the trace term, Eq.~\eqref{eqn:trace_ricci} can be recast in terms of the stress--energy tensor as
\begin{equation}
	\label{eqn:trace_stress_energy}
	\tr{\mathcal{H}} \LR = 8 \pi \, T_{a b} \, p^{a} p^{b} \LR .
\end{equation}

Since $p^{a} = ( \varepsilon / B ) ( p_{\phi} \xi^{a} + \eta^{a} )$ lies in the Killing two-plane, this may be written in coordinate form as
\begin{equation}
	\label{eqn:trace_components}
	\tr{\mathcal{H}} \LR = \frac{8 \pi}{B^{2}} ( p_{\phi}^{2} T_{t t} + 2 p_{\phi} T_{t \phi} + T_{\phi \phi} ) \LR ,
\end{equation}
so only the three independent components of $T_{a b}$ in the $(t , \phi)$-block enter. Note that, since the expression is quadratic in $p^{a}$, $\varepsilon$ disappears and the result is the same for prograde ($\varepsilon = 1$) and retrograde ($\varepsilon = -1$) light rings.

\subsubsection{Determinant}

By Eq.~\eqref{eqn:determinant_phi_psi}, all that remains is to identify the scalar $\Psi$. From Eqs.~\eqref{eqn:hessian_result} and \eqref{eqn:hessian_coefficients}, we have $\Psi = R_{a b c d} \, p^{a} m^{b} p^{c} m^{d}$ (after relabelling indices), in which the transverse projectors have been absorbed into the transverse vectors $m^{a}$. Decomposing the Riemann tensor into its Weyl and Ricci parts, every term arising from the Ricci part carries a metric factor pairing two of the vectors from the set $\{ m^{a} , m^{a} , p^{a} , p^{a} \}$. All such pairings vanish: $m^{a} m_{a} = 0$ by the normalisation \eqref{eqn:m_normalisation}, $p^{a} p_{a} = 0$ by the null condition, and $m^{a} p_{a} = 0$ because $m^{a}$ is transverse whilst $p^{a}$ lies in the Killing two-plane. The Ricci contribution therefore vanishes identically, leaving
\begin{equation}
	\label{eqn:psi_weyl}
	\Psi \LR = C_{a b c d} \, p^{a} m^{b} p^{c} m^{d} \LR = \Psi_{0} \LR ,
\end{equation}
where $\Psi_{0} = C_{a b c d} \, p^{a} m^{b} p^{c} m^{d}$ is one of the complex Weyl scalars of the Newman--Penrose formalism, defined with respect to the same tetrad, and $C_{a b c d}$ is the Weyl tensor.

The scalar $\Psi_{0}$ depends on the choice of $m^{a}$, but the residual freedom is only the spin rotation $m^{a} \mapsto e^{i \vartheta} m^{a}$, which preserves Eq.~\eqref{eqn:m_normalisation}. Under this transformation, $\Psi_{0} \mapsto e^{2 i \vartheta} \Psi_{0}$. The modulus $\lvert \Psi_{0} \rvert$, which is all that enters Eq.~\eqref{eqn:determinant_phi_psi}, is therefore invariant under this spin rotation. Combining Eqs.~\eqref{eqn:determinant_phi_psi}, \eqref{eqn:trace_ricci} and \eqref{eqn:psi_weyl}, the determinant of the transverse Hessian at a light ring is
\begin{equation}
	\label{eqn:determinant_result}
	\det{\mathcal{H}} \LR = ( \Phi_{0 0}^{2} - \lvert \Psi_{0} \rvert^{2} ) \LR .
\end{equation}

Equivalently, writing $\tilde{\mathcal{H}}_{a b} = \mathcal{H}_{a b} - \tfrac{1}{2} ( \tr{\mathcal{H}} ) P_{a b}$ for the trace-free part of the transverse Hessian, Eqs.~\eqref{eqn:determinant_phi_psi} and \eqref{eqn:psi_weyl} state that $\lvert \Psi_{0} \rvert^{2} = \tfrac{1}{2} \tilde{\mathcal{H}}_{a b} \tilde{\mathcal{H}}^{a b}$. The trace of the Hessian therefore carries the Ricci (focussing) content of the light ring, and its trace-free part the Weyl (tidal shear) content.

\subsubsection{Eigenvalues}

The two non-trivial eigenvalues of $\mathcal{H}^{a}{}_{b}$ now follow from $\tr{\mathcal{H}} = \mu_{+} + \mu_{-}$ and $\det{\mathcal{H}} = \mu_{+} \mu_{-}$. By Eqs.~\eqref{eqn:trace_ricci} and \eqref{eqn:determinant_result}, the discriminant of the resulting quadratic is a perfect square:
\begin{equation}
	\label{eqn:discriminant}
	[ ( \tr{\mathcal{H}} )^{2} - 4 \det{\mathcal{H}} ] \LR = 4 \lvert \Psi_{0} \rvert^{2} \LR .
\end{equation}
This confirms that both eigenvalues are real, as required for a real symmetric tensor. Hence
\begin{equation}
	\label{eqn:eigenvalues}
	\mu_{\pm} = ( \Phi_{0 0} \pm \lvert \Psi_{0} \rvert ) \LR
	= ( 4 \pi \, T_{a b} \, p^{a} p^{b} \pm \lvert \Psi_{0} \rvert ) \LR ,
\end{equation}
where the final equality uses the field equations as in Eq.~\eqref{eqn:trace_stress_energy}.

\subsection{Stability theorem}
\label{sec:stability_theorem}

We are now in a position to state the central result of this paper. The two stability criteria \eqref{eqn:stable_trace} and \eqref{eqn:stable_determinant}, which involve second derivatives of the metric functions in their original form, have been reduced to a pair of algebraic conditions on the eigenvalues \eqref{eqn:eigenvalues}. Since $\lvert \Psi_{0} \rvert \geq 0$, we have $\mu_{+} \geq \mu_{-}$ at any light ring, so the requirement that both eigenvalues be strictly positive is equivalent to the single requirement $\mu_{-} > 0$. The pair of stability criteria therefore collapses to one inequality.

\textbf{Theorem.} Let $(\mathcal{M}, g_{a b})$ be a four-dimensional circular stationary axisymmetric spacetime satisfying the Einstein field equations $G_{a b} = 8 \pi \, T_{a b}$ for a generic stress--energy tensor $T_{a b}$, and let $p^{a}$ be the tangent vector \eqref{eqn:tangent_killing} of a light ring situated off the rotation axis. Then the light ring is linearly stable if and only if
\begin{equation}
	\label{eqn:stability_theorem}
	 4 \pi \, T_{a b} \, p^{a} p^{b} \LR > \lvert \Psi_{0} \rvert \LR ,
\end{equation}
where $\Psi_{0} = C_{a b c d} \, p^{a} m^{b} p^{c} m^{d}$ is the Weyl scalar \eqref{eqn:psi_weyl} of the complex null tetrad $\{ p^{a} , n^{a} , m^{a} , \bar{m}^{a} \}$ adapted to the light ring.

The proof follows from our previous arguments. Stability requires $\mu_{\pm} > 0$, and by Eq.~\eqref{eqn:eigenvalues} the smaller eigenvalue is $\mu_{-} = ( 4 \pi \, T_{a b} \, p^{a} p^{b} - \lvert \Psi_{0} \rvert ) \LR$; the condition $\mu_{-} > 0$ is Eq.~\eqref{eqn:stability_theorem}, and it implies $\mu_{+} > 0$ automatically. Conversely, $\mu_{\pm} > 0$ gives Eq.~\eqref{eqn:stability_theorem} directly.

Purely in terms of spacetime curvature, without invoking the field equations, the condition \eqref{eqn:stability_theorem} is equivalent to
\begin{equation}
	\label{eqn:stability_theorem_curvature}
	\Phi_{0 0} \LR > \lvert \Psi_{0} \rvert \LR .
\end{equation}
By Eq.~\eqref{eqn:trace_ricci}, this may be written in terms of the Ricci tensor as $R_{a b} \, p^{a} p^{b} \LR > 2 \lvert \Psi_{0} \rvert \LR$. In components, using Eq.~\eqref{eqn:trace_components}, the same condition is
\begin{equation}
	\label{eqn:stability_theorem_components}
	\frac{4 \pi}{B^{2}} ( p_{\phi}^{2} T_{t t} + 2 p_{\phi} T_{t \phi} + T_{\phi \phi} ) \LR > \lvert \Psi_{0} \rvert \LR .
\end{equation}
The stability of a light ring is therefore determined entirely by two real scalars evaluated at a single point: the Ricci scalar $\Phi_{0 0}$, fixed by the matter content through the field equations, and the modulus of the Weyl scalar $\Psi_{0}$, which encodes vacuum curvature.

The two scalars enter with opposite effect. The Ricci term contributes equally to both eigenvalues, whilst the Weyl term raises one and lowers the other by the same amount; the Weyl curvature can therefore only ever act to destabilise a light ring. Stable light rings require the matter contribution to overcome the local tidal shear. Comparing $\Phi_{0 0} \LR$ and $\lvert \Psi_{0} \rvert \LR$, there are three important cases to consider.
\begin{itemize}
	\item If $\Phi_{0 0} > \lvert \Psi_{0} \rvert$, both eigenvalues are positive, the light ring is a local minimum of the geodesic potential $V$, and it is stable.
	\item If $\Phi_{0 0} < - \lvert \Psi_{0} \rvert$, both eigenvalues are negative, the light ring is a local maximum of $V$, and it is unstable.
	\item If $\lvert \Phi_{0 0} \rvert < \lvert \Psi_{0} \rvert$, the eigenvalues are of opposite sign, the light ring is a saddle point of $V$, and it is again unstable.
\end{itemize}
The two remaining possibilities $\Phi_{0 0} = \pm \lvert \Psi_{0} \rvert$ correspond to $\det{\mathcal{H}} \LR = 0$; these are the marginal cases excluded in Sec.~\ref{sec:transverse_hessian}, for which higher-order terms must be retained.

Although the normalisation $p_{t} = -1$ was fixed in Sec.~\ref{sec:hamiltonian_formalism} by an arbitrary rescaling of the affine parameter ($s \mapsto \alpha s$, $\alpha \in \mathbb{R}$), the condition \eqref{eqn:stability_theorem} is independent of that choice: both sides are quadratic in the tangent vector, so both scale as $\alpha^{-2}$ under $p^{a} \mapsto \alpha^{-1} p^{a}$, and the inequality is unaffected. Moreover, as noted below Eq.~\eqref{eqn:covariant_potential_2}, the sign $\varepsilon$ does not appear in Eq.~\eqref{eqn:stability_theorem}, so a single condition governs both the prograde and retrograde branches. However, the two light ring branches are associated with different values of $p_{\phi}$, and are generally located at different points, so Eq.~\eqref{eqn:stability_theorem} must be evaluated separately for each.

\section{Consequences of the stability theorem}
\label{sec:consequences}

\subsection{Equatorial light rings}
\label{sec:equatorial}

Many stationary axisymmetric spacetimes, including the Kerr solution, exhibit a $\mathbb{Z}_{2}$ (or ``north--south'') isometry. In adapted coordinates, this is a reflection symmetry about the equatorial plane $z = 0$: the metric is invariant under the transformation $z \mapsto -z$. In such spacetimes, the light rings often lie in the equatorial plane. This is the case for the rotating Kerr black hole, which admits two such light rings -- one prograde, the other retrograde \cite{Chandrasekhar1983}. However, the light rings of spacetimes with such a symmetry are not necessarily confined to the equatorial plane. For example, non-equatorial light rings have been uncovered (in mirror pairs) around the equal-mass Majumdar--Papapetrou binary black hole, which exhibits a reflective symmetry \cite{ShipleyDolan2016, Shipley2019}.

For the line element \eqref{eqn:sa_metric}, this condition means that the metric components $U$, $W$, $X$ and $\lambda$ are \emph{even} functions of $z$. As a result, the geodesic potential is also even in $z$, and the mixed component of the Hessian $\mathcal{H}_{\rho z}$ is odd in $z$. Hence, $\mathcal{H}_{\rho z} = 0$ at the equator $z = 0$.

Using the definition \eqref{eqn:hessian_coefficients} with $m^{a} \partial_{a} = ( e^{-\lambda} / \sqrt{2} ) ( \partial_{\rho} + i \, \partial_{z} )$, and the fact that $\Phi = \Phi_{0 0}$ and $\Psi = \Psi_{0 0}$ at a light ring, we may write the curvature scalars in adapted coordinates as $\Phi_{0 0} = \tfrac{1}{2} ( \mathcal{H}_{\rho \rho} + \mathcal{H}_{z z} )$ and $\Psi_{0} = \tfrac{1}{2} ( \mathcal{H}_{\rho \rho} - \mathcal{H}_{z z} ) + i \mathcal{H}_{\rho z}$ at the light ring. For an equatorial light ring, the imaginary term in the expression for $\Psi_{0}$ vanishes by the parity argument outlined above, so $\Psi_{0}$ is real at $z = 0$. The eigenvalues of the Hessian are then
\begin{equation}
	\label{eqn:eigenvalues_equatorial}
	\mu_{\rho} = ( \Phi_{0 0} + \Psi_{0} ) \LR , \qquad \mu_{z} = ( \Phi_{0 0} - \Psi_{0} ) \LR .
\end{equation}

For equatorial light rings in spacetimes with reflective symmetry, the stability condition \eqref{eqn:stability_theorem} is unchanged, but the sign of $\Psi_{0}$ now determines which of the radial and vertical directions is stable (and which is unstable) when the light ring corresponds to a saddle point of \eqref{eqn:potential_explicit}, i.e., in the case $\lvert \Phi_{0 0} \rvert < \lvert \Psi_{0} \rvert$. If $\Psi_{0} < 0$, an equatorial light ring is radially unstable but vertically stable, as for both light rings around a Kerr black hole. Conversely, if $\Psi_{0} > 0$, the light ring is radially stable but vertically unstable. Analysis of the radial potential alone would misclassify such a light ring as stable.

\subsection{Trace condition}
\label{sec:trace_condition}

The positivity of the trace alone provides a necessary criterion for stability, whose computation is inexpensive. Using Eq.~\eqref{eqn:trace_stress_energy}, we see that this is equivalent to the requirement $T_{a b} \, p^{a} p^{b} \LR > 0$, which is the statement that the \emph{strict} null energy condition holds at the light ring for the tangent vector $p^{a}$. Contrapositively, if the strict null energy condition is violated at a light ring, that light ring cannot be stable. This is a purely algebraic check involving the components of the stress--energy tensor in the $(t, \phi)$-block and the light ring's tangent vector, via Eq.~\eqref{eqn:trace_components}; it requires no knowledge of the Weyl curvature.

The condition is necessary but not sufficient. A light ring at which the strict null energy condition holds is nevertheless a saddle point of the geodesic potential, and hence unstable, whenever $\lvert \Psi_{0} \rvert$ exceeds $\Phi_{0 0}$ there; the Weyl curvature acts only ever to destabilise. Sufficiency is recovered when $\Psi_{0} \LR = 0$. This occurs when $p^{a}$ is a principal null direction of the Weyl tensor at the light ring, in agreement with the corresponding result of Ref.~\cite{KogaHarada2019} for photon surfaces (Corollary 5). The relationship to the null energy condition result of Ref.~\cite{CunhaBertiHerdeiro2017}, which is sufficient given the topological identification of a light ring as a non-saddle critical point, is discussed in Sec.~\ref{sec:discussion}.

We note that the condition \eqref{eqn:stability_theorem} is linear in the stress--energy tensor. If the matter content is a sum of non-interacting sectors, $T_{a b} = \sum_{i} T^{(i)}{}_{a b}$, then the contributions to $\tr{\mathcal{H}}$ simply add. Consequently, a matter sector that contributes nothing to $T_{a b} \, p^{a} p^{b}$ at a light ring can neither create nor destroy stability.

Relatedly, any contribution to the stress--energy tensor proportional to the metric is annihilated. If $T^{(i)}{}_{a b} = K \, g_{a b}$ for some scalar $K$, then $T^{(i)}{}_{a b} \, p^{a} p^{b} \LR = K \, p^{a} p_{a} \LR = 0$ by the null condition. This mechanism removes every term proportional to $g_{a b}$ in the matter models considered below, and it is responsible for the first of our results.

\subsection{Application to matter models}
\label{sec:matter_models}

We now apply the stability theorem to a selection of stress--energy tensors of physical interest. In each case the task reduces to evaluating the double null contraction $T_{a b} \, p^{a} p^{b}$ at the light ring, where $p^{a}$ is the tangent vector \eqref{eqn:tangent_killing}. Two properties of $p^{a}$ play a key role: it is null, which removes every term proportional to $g_{a b}$; and it lies in the Killing two-plane, which constrains how the matter fields may vary along the ray. Throughout, we assume that the matter fields are compatible with the circularity conditions \eqref{eqn:circularity_conditions}, in the form of the component conditions \eqref{eqn:circularity_components}.

\subsubsection{Vacuum with a cosmological constant}

In vacuum with a cosmological constant $\Lambda$, the cosmological term $\Lambda g_{a b}$ can be moved to the right-hand side of Einstein's field equations, and therefore viewed as a source of vacuum energy, with effective stress--energy tensor
\begin{equation}
	\label{eqn:T_lambda}
	T_{a b} = - \frac{\Lambda}{8 \pi} g_{a b} .
\end{equation}
As this is proportional to the metric tensor, the null contraction yields $T_{a b} \, p^{a} p^{b} \LR = 0$ identically. Hence $\tr{\mathcal{H}} \LR = 0$, or equivalently $\Phi_{0 0} \LR = 0$, and the eigenvalues \eqref{eqn:eigenvalues} reduce to
\begin{equation}
	\label{eqn:eigenvalues_lambda}
	\mu_{\pm} = \pm \lvert \Psi_{0} \rvert \LR ,
\end{equation}
with determinant
\begin{equation}
	\label{eqn:det_lambda}
	\det{\mathcal{H}} \LR = - \lvert \Psi_{0} \rvert^{2} \LR \leq 0 .
\end{equation}
The determinant condition \eqref{eqn:stable_determinant} therefore fails for any $\Psi_{0} \neq 0$: the light ring is a saddle point of the geodesic potential and is unstable. In the degenerate case $\Psi_{0} \LR = 0$ both eigenvalues vanish and the light ring is marginal, lying outside the scope of the linear analysis. Stable light rings are thus ruled out in vacuum with a cosmological constant, recovering the central result of Ref.~\cite{Shipley2023}; the case of pure vacuum, explored in Ref.~\cite{DolanShipley2016}, follows on setting $\Lambda = 0$.

For equatorial light rings in vacuum spacetimes with $\mathbb{Z}_{2}$ reflective symmetry, the eigenvalues are $\mu_{\rho} = \Psi_{0} \LR$ and $\mu_{z} = - \Psi_{0} \LR$, with $\Psi_{0}$ real; see Sec.~\ref{sec:equatorial}. This result applies to the rotating Kerr family, which includes the static Schwarzschild solution as a special case. The Kerr black hole admits a pair of equatorial light rings: an outer retrograde (counter-rotating) orbit, and an inner prograde (co-rotating) orbit, both of which are unstable. For the non-rotating Schwarzschild case, these two light rings coincide: there is a single unstable light ring (which may be traversed in either rotation sense).

Starting with the Weyl form of the Schwarzschild metric for a black hole of mass $M$ \cite{Weyl1917, GriffithsPodolsky2009}, it is straightforward to show using the conditions \eqref{eqn:covariant_light_ring_conditions} that the light ring is located at $\rho = \sqrt{3} M$, $z = 0$, with $p_{\phi} = 1/\Omega = \pm 3 \sqrt{3} M$. (Using the relationship between the Weyl coordinates and more familiar Schwarzschild coordinates, one may show that the light ring is situated at $r = 3 M$, $\theta = \pi/2$.) The Schwarzschild spacetime is a vacuum solution ($\Phi_{0 0} = 0$), and the light ring is equatorial, so the eigenvalues of the Hessian take the form $\mu_{\rho} = \Psi_{0} \LR$ and $\mu_{z} = - \Psi_{0} \LR$, where $\Psi_{0}$ is real, as described above. Hence, the stability of the light ring is fully determined by the Weyl scalar $\Psi_{0}$ evaluated at the light ring. With tetrad legs $p^{a} \partial_{a} = 3 \, \partial_{t} + \varepsilon \sqrt{3} / (3M) \, \partial_{\phi}$ by the identity \eqref{eqn:tangent_killing} and $m^{a} \partial_{a} = ( e^{-\lambda} / \sqrt{2} ) ( \partial_{\rho} + i \, \partial_{z} )$, direct computation of $\Psi_{0} = C_{a b c d}  \, p^{a} m^{b} p^{c} m^{d}$ (using a symbolic algebra package) yields $\Psi_{0} \LR = - 1 / (3 M^{2})$, which is real, as expected. Hence $\mu_{\rho} = - 1 / (3 M^{2}) < 0$ and $\mu_{z} = 1 / (3 M^{2}) > 0$, so the light ring corresponds to a saddle, which is radially unstable and vertically stable, as is well-known for the Schwarzschild black hole.

\subsubsection{Electrovacuum}

For a source-free electromagnetic field with field strength $\mathcal{F}_{a b}$, not to be confused with the Killing bivector $F_{a b}$ of Eq.~\eqref{eqn:killing_bivector_def}, the stress--energy tensor is
\begin{equation}
	\label{eqn:T_electrovacuum}
	T_{a b} = \tfrac{1}{4 \pi} \left( \mathcal{F}_{a c} \mathcal{F}_{b}{}^{c} - \tfrac{1}{4} g_{a b} \mathcal{F}_{c d} \mathcal{F}^{c d} \right) .
\end{equation}
The second term is annihilated by the null condition, and introducing the covector
\begin{equation}
	\label{eqn:photon_electric_field}
	\mathcal{E}_{a} = \mathcal{F}_{a b} \, p^{b} ,
\end{equation}
which is the electric field along the light ring, the double null contraction becomes
\begin{equation}
	\label{eqn:T_electrovacuum_contraction}
	4 \pi \, T_{a b} \, p^{a} p^{b} \LR = \mathcal{E}_{a} \mathcal{E}^{a} \LR .
\end{equation}
The antisymmetry of $\mathcal{F}_{a b}$ gives $\mathcal{E}_{a} p^{a} = 0$, so $\mathcal{E}_{a}$ is orthogonal to the null vector $p^{a}$ and is therefore either spacelike or proportional to $p_{a}$ itself. Consequently $\mathcal{E}_{a} \mathcal{E}^{a} \geq 0$: the null energy condition is satisfied in electrovacuum, and the stability condition \eqref{eqn:stability_theorem} reads
\begin{equation}
	\label{eqn:stability_electrovacuum}
	\mathcal{E}_{a} \mathcal{E}^{a} \LR > \lvert \Psi_{0} \rvert \LR .
\end{equation}
Generic stable light rings are therefore permitted in electrovacuum, in agreement with Refs.~\cite{DolanShipley2016, Shipley2023}, provided the electromagnetic field at the light ring is strong enough to overcome the local Weyl curvature. They are excluded in the degenerate case $\mathcal{E}_{a} \propto p_{a}$, in which the electromagnetic contribution to the trace vanishes and Eq.~\eqref{eqn:eigenvalues_lambda} applies.

By the linearity property noted in Sec.~\ref{sec:transverse_hessian}, the addition of a cosmological constant to an electrovacuum spacetime alters neither side of Eq.~\eqref{eqn:stability_electrovacuum}, since the contribution \eqref{eqn:T_lambda} is annihilated. This explains why the introduction of $\Lambda$ could not forbid stable light rings in the electrovacuum spacetimes examined in Ref.~\cite{Shipley2023}.

In Refs.~\cite{DolanShipley2016, Shipley2023} we noted that a ``remarkable cancellation'' occurred on substituting the field equations into the expression for the Hessian, whereby the derivatives of the metric functions dropped out and only stress--energy terms remained. The present analysis identifies the origin of that cancellation: as established in Sec.~\ref{sec:curvature_identities}, the Killing bivector terms of Eq.~\eqref{eqn:covariant_potential_2} are annihilated exactly by the transverse projectors at a light ring, leaving only curvature terms, which can be exchanged for matter terms via the Einstein field equations. This cancellation is therefore not only a property of electrovacuum spacetimes with a cosmological constant, but a generic feature of circular stationary axisymmetric geometries.

\subsubsection{Real scalar field}

For a minimally coupled real scalar field $\varphi$ with self-interaction potential $\mathcal{V}(\varphi)$, the stress--energy tensor is
\begin{equation}
	\label{eqn:T_real_scalar}
	T_{a b} = \nabla_{a} \varphi \, \nabla_{b} \varphi - g_{a b} \left( \tfrac{1}{2} \nabla_{c} \varphi  \, \nabla^{c} \varphi + \mathcal{V}( \varphi ) \right) .
\end{equation}
For a general null vector $p^{a}$, the double contraction of the second term vanishes, and we are left with
\begin{equation}
	\label{eqn:T_real_scalar_contraction}
	T_{a b} \, p^{a} p^{b} = ( p^{a} \nabla_{a} \varphi )^{2} \geq 0 .
\end{equation}
The null energy condition is therefore satisfied, but not, in general, strictly.

In adapted coordinates, we must have $\varphi = \varphi(\rho, z)$, since otherwise the stress--energy tensor for a real scalar field \eqref{eqn:T_real_scalar} would fail to be stationary and axisymmetric. This means that $\nabla_{a} \varphi$ lies in the transverse two-plane. As the tangent vector at a light ring lies in the Killing two-plane, it is orthogonal to $\nabla_{a} \varphi$, i.e., $p^{a} \nabla_{a} \varphi \LR = 0$. Hence, the inequality \eqref{eqn:T_real_scalar_contraction} is saturated at the light ring, so the eigenvalues reduce to the form \eqref{eqn:eigenvalues_lambda} obtained in vacuum, and $\det{\mathcal{H}} \LR \leq 0$.

First-order stable light rings are therefore ruled out for a minimally coupled real scalar field, for \emph{any} self-interaction potential $\mathcal{V}(\varphi)$. The mechanism differs from that of the cosmological constant: in that case, the contraction vanishes because $p^{a}$ is null; here, it vanishes because $p^{a}$ generates an isometry. To our knowledge, this result has not previously been presented in the literature.

\subsubsection{Complex scalar field}

The situation changes for a complex scalar field $\psi$, with stress--energy tensor
\begin{equation}
	\label{eqn:T_complex_scalar}
	T_{a b} = \nabla_{a} \psi \, \nabla_{b} \bar{\psi} + \nabla_{a} \bar{\psi} \, \nabla_{b} \psi - g_{a b} \left( \nabla_{c} \psi  \, \nabla^{c} \bar{\psi} + \mathcal{V}( \lvert \psi \rvert^{2} ) \right) ,
\end{equation}
for which the double null contraction gives
\begin{equation}
	\label{eqn:T_complex_scalar_contraction}
	T_{a b} \, p^{a} p^{b} = 2 \lvert p^{a} \nabla_{a} \psi \rvert^{2} \geq 0 .
\end{equation}
for a generic null vector $p^{a}$. Again, the stress--energy term proportional to $g_{a b}$ is killed by the null condition. However, the argument employed for the real scalar field no longer applies, because $T_{a b}$ depends on $\psi$ only through the invariant combinations $\lvert \psi \rvert^{2}$ and $\nabla_{a} \bar{\psi} \nabla_{b} \psi$. The scalar field itself may therefore carry harmonic dependence on the Killing coordinates $t$ and $\phi$, whilst leaving the stress--energy tensor stationary and axisymmetric, as in the ansatz \cite{YoshidaEriguchi1997, KleihausKunzList2005, Grandclement2017}
\begin{equation}
	\label{eqn:harmonic_ansatz}
	\psi = A( \rho, z ) \, e^{i ( m \phi - \omega t )} ,
\end{equation}
with real amplitude $A$, frequency $\omega$ and azimuthal harmonic index $m \in \mathbb{Z}$ (where the latter should not be confused with the complex null tetrad leg $m^{a}$). Substituting Eq.~\eqref{eqn:harmonic_ansatz} and using $p^{a} \nabla_{a} \psi = ( \varepsilon / B ) ( p_{\phi} \, \partial_{t} + \partial_{\phi} ) \, \psi$ at the light ring, we obtain
\begin{equation}
	\label{eqn:stability_complex_scalar}
	4 \pi \, T_{a b} \, p^{a} p^{b} \LR = \frac{8 \pi \, A^{2}}{B^{2}} ( m - \omega \, p_{\phi} )^{2} \LR \geq 0 .
\end{equation}
The right-hand side is strictly positive, except on the synchronisation condition $\omega = m \, \Omega$, where $\Omega = 1 / p_{\phi}$ is the angular velocity of the light ring. A complex scalar field can therefore support stable light rings, provided the right-hand side of Eq.~\eqref{eqn:stability_complex_scalar} exceeds $\lvert \Psi_{0} \rvert$ there.

Contrasting with the result for the real scalar field, which is recovered from Eq.~\eqref{eqn:stability_complex_scalar} on setting $\omega = 0 = m$, we see that it is not the scalar sector that forbids stable light rings, but the requirement that a real field be invariant under the isometries, which forces the field to be constant along the light ring.

Minimally coupled complex scalar fields have been used to construct horizonless, self-gravitating objects known as Boson stars \cite{Kaup1968}, both in their static spherically symmetric form \cite{RuffiniBonazzola1969} and in their stationary axisymmetric form \cite{YoshidaEriguchi1997, SchunckMielke2003}, as well generalisations with a horizon in the form of Kerr black holes with scalar hair \cite{HerdeiroRadu2014}. The circular photon orbits of these spacetimes have been studied in detail, and stable light rings have been revealed \cite{CunhaEtAl2015, CunhaEtAl2016, CunhaBertiHerdeiro2017, CunhaEtAl2017, Grandclement2017}, in agreement with the result presented above.

Complex Proca (massive vector) fields provide an analogous matter model. These have been used to build Proca stars \cite{BritoEtAl2016} and Kerr black holes with Proca hair \cite{HerdeiroRaduRunarsson2016}. As with the complex scalar case, stable light rings have been identified around ultra-compact Proca stars and black holes with Proca hair \cite{SengoEtAl2023}. Our key result -- particularly the trace test -- could be applied to the stress--energy tensor for the complex Proca field, using the approach laid out above for vacuum, electrovacuum and scalar fields. This is left as an extension, along with the application to other matter models of interest.

\section{Discussion}
\label{sec:discussion}

In this work, we have obtained a necessary and sufficient condition for the stability of a light ring in a four-dimensional circular stationary axisymmetric spacetime with generic stress--energy, Eq.~\eqref{eqn:stability_theorem}, and applied it to a range of matter models.

Our result is local and geometric, as shown by the form of Eq.~\eqref{eqn:stability_theorem_curvature}, and complements the global topological arguments of Refs.~\cite{CunhaBertiHerdeiro2017, CunhaHerdeiro2020, GhoshSarkar2021, CunhaHerdeiroNovo2024}, which establish the existence and pairing of light rings. Here, we assume that a light ring exists and say nothing about how many there are; conversely, the topological arguments focus on existence and say nothing about the stability of a given light ring in isolation.

The point of closest contact with the present study is the second part of the theorem of Cunha, Berti and Herdeiro \cite{CunhaBertiHerdeiro2017}: the null energy condition renders the non-saddle member of a light ring pair a minimum of the geodesic potential. The authors also derive the trace of the Hessian of the geodesic potential in terms of $T_{a b} p^{a} p^{b}$ at a light ring, via direct computation of the Christoffel symbols. The derivation given here instead writes the second derivatives of the metric functions as quantities bilinear in the Killing vectors, Eqs.~\eqref{eqn:U_bilinear}--\eqref{eqn:W_bilinear}, and the field-strength terms that arise in Eq.~\eqref{eqn:covariant_potential_2} are annihilated by the transverse projector at a light ring, which makes clear the role of the spacetime's Killing symmetries. We also obtain the determinant \eqref{eqn:determinant_result}, which is needed for a complete classification, and which permits us to obtain the eigenvalues in Eq.~\eqref{eqn:eigenvalues}. The condition \eqref{eqn:stability_theorem} is moreover pointwise: it applies to a single light ring, with no topological input identifying it as a non-saddle critical point. The conclusion of Ref.~\cite{CunhaBertiHerdeiro2017} follows on noting that the strict null energy condition gives $\tr{\mathcal{H}} \LR > 0$, so that a critical point already known to be an extremum must be a minimum. Conversely, Eq.~\eqref{eqn:eigenvalues} shows that the null energy condition alone is not sufficient for stability: a light ring at which $\lvert \Psi_{0} \rvert$ exceeds $\Phi_{0 0}$ is a saddle even for conventional matter. We also note that the expression for the trace derived in Ref.~\cite{CunhaBertiHerdeiro2017} differs from Eq.~\eqref{eqn:trace_stress_energy} by a factor of two, which is accounted for by the fact that the geodesic potential in that work is defined as twice ours.

Koga and Harada \cite{KogaHarada2019} obtain a stability criterion for null geodesics on a photon \emph{surface} \cite{ClaudelVirbhadraEllis2001, Perlick2005} from the geodesic deviation equation, in terms of the Riemann tensor contracted twice with the null tangent and twice with the unit normal of the surface. Their criterion is the contraction of our transverse Hessian \eqref{eqn:hessian_result} along a single transverse direction, and their Weyl--Ricci decomposition reproduces, in four dimensions, the Ricci contribution $\Phi_{0 0}$ of Eq.~\eqref{eqn:trace_ricci}. Several of their corollaries for photon surfaces follow for light rings from Eq.~\eqref{eqn:eigenvalues}. In particular, if the spacetime has constant curvature, or is conformally flat, or if the tangent vector is a principal null direction of the Weyl tensor, then $\Psi_{0} \LR = 0$, so $\mu_{\pm} = \Phi_{0 0} \LR$. The light ring is therefore marginal in the case of constant-curvature spacetimes ($\Phi_{00} = 0$). Assuming the null energy condition ($\Phi_{00} \geq 0$), a light ring may be stable or marginal in the other two cases.

Despite these connections, the present study is complementary to Ref.~\cite{KogaHarada2019}. A photon surface is a hypersurface, so its unit normal defines a single perturbation direction, and the stability considered in Ref.~\cite{KogaHarada2019} is that of a null geodesic with respect to the surface containing it; perturbations tangent to the surface are not considered, by the defining property of a photon surface. A light ring is a curve, and the transverse plane is two-dimensional, so the linearised stability analysis requires the full rank-two Hessian \eqref{eqn:transverse_hessian} and both of its invariants. Moreover, the underlying assumptions are not nested: Ref.~\cite{KogaHarada2019} requires a totally umbilic hypersurface, and we require stationarity, axisymmetry and circularity.

As shown in Sec.~\ref{sec:scalar_invariants}, one may construct a complex null tetrad at a light ring, and the trace and trace-free parts of the transverse Hessian are, respectively, the Ricci ($\Phi_{00}$) and Weyl ($\Psi_{0}$) scalars of that tetrad. These are precisely the quantities that feature in the Sachs optical equations \cite{Sachs1962}, which read
\begin{align}
	\dot{\varrho} &= \varrho^{2} + \sigma \bar{\sigma} + \Phi_{0 0} , \\
	\dot{\sigma} &= \sigma ( \varrho + \bar{\varrho} ) + \Psi_{0} ,
\end{align}
where $\varrho$ and $\sigma$ are Newman--Penrose scalars (also referred to as the \emph{optical scalars}), with $\operatorname{Re}{(\varrho)}$ describing the expansion, $\operatorname{Im}{(\varrho)}$ the twist, and $\sigma$ the shear of a null geodesic congruence. It is therefore conceivable that the results of the present work may be derived from the geodesic deviation equation, using the Newman--Penrose formalism, with stability characterised by the focussing and shear of a congruence centred on the light ring. Indeed, Ref.~\cite{KogaHarada2019} proceeds in that way for the study of photon surfaces and their stability. Here, we prefer the Hamiltonian approach because the role of the spacetime's Killing symmetries is manifest, and it delivers the full two-dimensional transverse Hessian.

Qiao \emph{et al.}~\cite{Qiao2025} employ a geometric approach to study equatorial light rings in stationary axisymmetric spacetimes, recasting them in terms of a Randers--Finsler geometry. They demonstrate that light rings can be determined by the vanishing of geodesic curvature, and their stability classified using the intrinsic flag curvature in the Randers--Finsler optical geometry. The geometric approach presented in that work is applicable to any stationary and axially symmetric spacetime, and the authors demonstrate that it is equivalent to the conventional effective potential method.

The (trace-reversed) field equations enter only through the substitution to derive Eqs.~\eqref{eqn:trace_stress_energy}, so the stability theorem stated in Eq.~\eqref{eqn:stability_theorem} holds for any modified theory of gravity in which photons propagate along null geodesics, with $T_{a b}$ replaced by the effective stress--energy tensor of that theory. The geometric form \eqref{eqn:stability_theorem_curvature} requires no field equations at all.

We have imposed circularity throughout this work, which is automatic in electrovacuum \cite{Carter1969, KundtTrumper1966} but is an additional hypothesis for generic matter. Relaxing it would destroy the block-diagonal form of the metric \eqref{eqn:sa_metric}, and with it the Hamiltonian formalism and the geodesic potential \eqref{eqn:potential_explicit}, as well as the effective potentials \eqref{eqn:effective_potentials} used in Refs.~\cite{DolanShipley2016, Shipley2019, Shipley2023} and elsewhere. The non-circular case (retaining stationarity and axisymmetry) would be a natural generalisation of this work, as would the extension to higher dimensions, where the transverse space has dimension greater than two \cite{IgataTomizawa2020, TavlayanTekin2023}.

Beyond circular orbits, recent studies have looked at so-called \emph{fundamental photon orbits} \cite{ShipleyDolan2016, CunhaHerdeiroRadu2017, LimaJuniorEtAl2021}, which can be thought of as generalisations of the spherical photon orbits of Kerr spacetime \cite{Teo2003} to stationary axisymmetric spacetimes on which the geodesic equations are not necessarily integrable. These more general fundamental photon orbits fall beyond the stability analysis presented in this work, which applies only to light rings. In Ref.~\cite{CunhaHerdeiroRadu2017}, a classification scheme for fundamental photon orbits was introduced, and their stability analysed using Poincar\'{e} maps. Further, it was demonstrated that around Kerr black holes with Proca hair, \emph{stable} fundamental photon orbits exist, even when the light rings are unstable. Like stable light rings, stable fundamental photon orbits are associated with spacetime instabilities, and are responsible for novel features observed in black hole shadows and lensing patterns. Extending the stability analysis of light rings covered in the present paper to fundamental photon orbits would be of interest, and would provide a more complete description of geodesic motion and gravitational lensing phenomena in four-dimensional stationary axisymmetric geometries.

\begin{acknowledgments}
The author thanks Sam Dolan for helpful discussions and feedback. During the preparation of this paper, the author used generative artificial intelligence (AI) tools, namely Claude Opus 5, to generate research ideas, perform literature search, and edit text. The author has reviewed and validated all AI-assisted outputs, and takes full responsibility for the content of the manuscript.
\end{acknowledgments}

\bibliographystyle{apsrev4-2}
\bibliography{light_ring_stability_conditions_refs}

\begin{thebibliography}{58}%
\makeatletter
\providecommand \@ifxundefined [1]{%
 \@ifx{#1\undefined}
}%
\providecommand \@ifnum [1]{%
 \ifnum #1\expandafter \@firstoftwo
 \else \expandafter \@secondoftwo
 \fi
}%
\providecommand \@ifx [1]{%
 \ifx #1\expandafter \@firstoftwo
 \else \expandafter \@secondoftwo
 \fi
}%
\providecommand \natexlab [1]{#1}%
\providecommand \enquote  [1]{``#1''}%
\providecommand \bibnamefont  [1]{#1}%
\providecommand \bibfnamefont [1]{#1}%
\providecommand \citenamefont [1]{#1}%
\providecommand \href@noop [0]{\@secondoftwo}%
\providecommand \href [0]{\begingroup \@sanitize@url \@href}%
\providecommand \@href[1]{\@@startlink{#1}\@@href}%
\providecommand \@@href[1]{\endgroup#1\@@endlink}%
\providecommand \@sanitize@url [0]{\catcode `\\12\catcode `\$12\catcode
  `\&12\catcode `\#12\catcode `\^12\catcode `\_12\catcode `\%12\relax}%
\providecommand \@@startlink[1]{}%
\providecommand \@@endlink[0]{}%
\providecommand \url  [0]{\begingroup\@sanitize@url \@url }%
\providecommand \@url [1]{\endgroup\@href {#1}{\urlprefix }}%
\providecommand \urlprefix  [0]{URL }%
\providecommand \Eprint [0]{\href }%
\providecommand \doibase [0]{https://doi.org/}%
\providecommand \selectlanguage [0]{\@gobble}%
\providecommand \bibinfo  [0]{\@secondoftwo}%
\providecommand \bibfield  [0]{\@secondoftwo}%
\providecommand \translation [1]{[#1]}%
\providecommand \BibitemOpen [0]{}%
\providecommand \bibitemStop [0]{}%
\providecommand \bibitemNoStop [0]{.\EOS\space}%
\providecommand \EOS [0]{\spacefactor3000\relax}%
\providecommand \BibitemShut  [1]{\csname bibitem#1\endcsname}%
\let\auto@bib@innerbib\@empty
%</preamble>
\bibitem [{\citenamefont {Dyson}\ \emph {et~al.}(1920)\citenamefont {Dyson},
  \citenamefont {Eddington},\ and\ \citenamefont
  {Davidson}}]{DysonEddingtonDavidson1920}%
  \BibitemOpen
  \bibfield  {author} {\bibinfo {author} {\bibfnamefont {F.~W.}\ \bibnamefont
  {Dyson}}, \bibinfo {author} {\bibfnamefont {A.~S.}\ \bibnamefont
  {Eddington}},\ and\ \bibinfo {author} {\bibfnamefont {C.}~\bibnamefont
  {Davidson}},\ }\href {https://doi.org/10.1098/rsta.1920.0009} {\bibfield
  {journal} {\bibinfo  {journal} {Phil. Trans. Roy. Soc. Lond. A}\ }\textbf
  {\bibinfo {volume} {220}},\ \bibinfo {pages} {291} (\bibinfo {year}
  {1920})}\BibitemShut {NoStop}%
\bibitem [{\citenamefont {Wambsganss}(1998)}]{Wambsganss1998}%
  \BibitemOpen
  \bibfield  {author} {\bibinfo {author} {\bibfnamefont {J.}~\bibnamefont
  {Wambsganss}},\ }\href {https://doi.org/10.12942/lrr-1998-12} {\bibfield
  {journal} {\bibinfo  {journal} {Living Rev. Rel.}\ }\textbf {\bibinfo
  {volume} {1}},\ \bibinfo {pages} {12} (\bibinfo {year} {1998})},\ \Eprint
  {https://arxiv.org/abs/astro-ph/9812021} {arXiv:astro-ph/9812021}
  \BibitemShut {NoStop}%
\bibitem [{\citenamefont {Perlick}(2004)}]{Perlick2004}%
  \BibitemOpen
  \bibfield  {author} {\bibinfo {author} {\bibfnamefont {V.}~\bibnamefont
  {Perlick}},\ }\href {https://doi.org/10.12942/lrr-2004-9} {\bibfield
  {journal} {\bibinfo  {journal} {Living Rev. Rel.}\ }\textbf {\bibinfo
  {volume} {7}},\ \bibinfo {pages} {1} (\bibinfo {year} {2004})},\ \Eprint
  {https://arxiv.org/abs/1010.3416} {arXiv:1010.3416 [gr-qc]} \BibitemShut
  {NoStop}%
\bibitem [{\citenamefont {Cunha}\ and\ \citenamefont
  {Herdeiro}(2018)}]{CunhaHerdeiro2018}%
  \BibitemOpen
  \bibfield  {author} {\bibinfo {author} {\bibfnamefont {P.~V.~P.}\
  \bibnamefont {Cunha}}\ and\ \bibinfo {author} {\bibfnamefont {C.~A.~R.}\
  \bibnamefont {Herdeiro}},\ }\href {https://doi.org/10.1007/s10714-018-2361-9}
  {\bibfield  {journal} {\bibinfo  {journal} {{Gen. Relativ. Gravitation}}\
  }\textbf {\bibinfo {volume} {50}},\ \bibinfo {pages} {1} (\bibinfo {year}
  {2018})},\ \Eprint {https://arxiv.org/abs/1801.00860} {arXiv:1801.00860
  [gr-qc]} \BibitemShut {NoStop}%
\bibitem [{\citenamefont {Synge}(1966)}]{Synge1966}%
  \BibitemOpen
  \bibfield  {author} {\bibinfo {author} {\bibfnamefont {J.~L.}\ \bibnamefont
  {Synge}},\ }\href {https://doi.org/10.1093/mnras/131.3.463} {\bibfield
  {journal} {\bibinfo  {journal} {Mon. Not. Roy. Astron. Soc.}\ }\textbf
  {\bibinfo {volume} {131}},\ \bibinfo {pages} {463} (\bibinfo {year}
  {1966})}\BibitemShut {NoStop}%
\bibitem [{\citenamefont {Akiyama}\ \emph {et~al.}(2019)\citenamefont {Akiyama}
  \emph {et~al.}}]{EventHorizonTelescope2019}%
  \BibitemOpen
  \bibfield  {author} {\bibinfo {author} {\bibfnamefont {K.}~\bibnamefont
  {Akiyama}} \emph {et~al.} (\bibinfo {collaboration} {Event Horizon
  Telescope}),\ }\href {https://doi.org/10.3847/2041-8213/ab0ec7} {\bibfield
  {journal} {\bibinfo  {journal} {Astrophys. J. Lett.}\ }\textbf {\bibinfo
  {volume} {875}},\ \bibinfo {pages} {L1} (\bibinfo {year} {2019})},\ \Eprint
  {https://arxiv.org/abs/1906.11238} {arXiv:1906.11238 [astro-ph.GA]}
  \BibitemShut {NoStop}%
\bibitem [{\citenamefont {Cunha}\ \emph {et~al.}(2016)\citenamefont {Cunha},
  \citenamefont {Grover}, \citenamefont {Herdeiro}, \citenamefont {Radu},
  \citenamefont {R\'{u}narsson},\ and\ \citenamefont {Wittig}}]{CunhaEtAl2016}%
  \BibitemOpen
  \bibfield  {author} {\bibinfo {author} {\bibfnamefont {P.~V.~P.}\
  \bibnamefont {Cunha}}, \bibinfo {author} {\bibfnamefont {J.}~\bibnamefont
  {Grover}}, \bibinfo {author} {\bibfnamefont {C.}~\bibnamefont {Herdeiro}},
  \bibinfo {author} {\bibfnamefont {E.}~\bibnamefont {Radu}}, \bibinfo {author}
  {\bibfnamefont {H.}~\bibnamefont {R\'{u}narsson}},\ and\ \bibinfo {author}
  {\bibfnamefont {A.}~\bibnamefont {Wittig}},\ }\href
  {https://doi.org/10.1103/PhysRevD.94.104023} {\bibfield  {journal} {\bibinfo
  {journal} {{Phys. Rev. D}}\ }\textbf {\bibinfo {volume} {94}},\ \bibinfo
  {pages} {104023} (\bibinfo {year} {2016})},\ \Eprint
  {https://arxiv.org/abs/1609.01340} {arXiv:1609.01340 [gr-qc]} \BibitemShut
  {NoStop}%
\bibitem [{\citenamefont {Cunha}\ \emph
  {et~al.}(2017{\natexlab{a}})\citenamefont {Cunha}, \citenamefont {Font},
  \citenamefont {Herdeiro}, \citenamefont {Radu}, \citenamefont
  {Sanchis-Gual},\ and\ \citenamefont {Zilh{\~a}o}}]{CunhaEtAl2017}%
  \BibitemOpen
  \bibfield  {author} {\bibinfo {author} {\bibfnamefont {P.~V.~P.}\
  \bibnamefont {Cunha}}, \bibinfo {author} {\bibfnamefont {J.~A.}\ \bibnamefont
  {Font}}, \bibinfo {author} {\bibfnamefont {C.}~\bibnamefont {Herdeiro}},
  \bibinfo {author} {\bibfnamefont {E.}~\bibnamefont {Radu}}, \bibinfo {author}
  {\bibfnamefont {N.}~\bibnamefont {Sanchis-Gual}},\ and\ \bibinfo {author}
  {\bibfnamefont {M.}~\bibnamefont {Zilh{\~a}o}},\ }\href
  {https://doi.org/10.1103/PhysRevD.96.104040} {\bibfield  {journal} {\bibinfo
  {journal} {Phys. Rev. D}\ }\textbf {\bibinfo {volume} {96}},\ \bibinfo
  {pages} {104040} (\bibinfo {year} {2017}{\natexlab{a}})},\ \Eprint
  {https://arxiv.org/abs/1709.06118} {arXiv:1709.06118 [gr-qc]} \BibitemShut
  {NoStop}%
\bibitem [{\citenamefont {Cardoso}\ \emph {et~al.}(2016)\citenamefont
  {Cardoso}, \citenamefont {Franzin},\ and\ \citenamefont
  {Pani}}]{CardosoFranzinPani2016}%
  \BibitemOpen
  \bibfield  {author} {\bibinfo {author} {\bibfnamefont {V.}~\bibnamefont
  {Cardoso}}, \bibinfo {author} {\bibfnamefont {E.}~\bibnamefont {Franzin}},\
  and\ \bibinfo {author} {\bibfnamefont {P.}~\bibnamefont {Pani}},\ }\href
  {https://doi.org/10.1103/PhysRevLett.116.171101} {\bibfield  {journal}
  {\bibinfo  {journal} {Phys. Rev. Lett.}\ }\textbf {\bibinfo {volume} {116}},\
  \bibinfo {pages} {171101} (\bibinfo {year} {2016})},\ \Eprint
  {https://arxiv.org/abs/1602.07309} {arXiv:1602.07309 [gr-qc]} \BibitemShut
  {NoStop}%
\bibitem [{\citenamefont {Keir}(2016)}]{Keir2016}%
  \BibitemOpen
  \bibfield  {author} {\bibinfo {author} {\bibfnamefont {J.}~\bibnamefont
  {Keir}},\ }\href {https://doi.org/10.1088/0264-9381/33/13/135009} {\bibfield
  {journal} {\bibinfo  {journal} {Class. Quant. Grav.}\ }\textbf {\bibinfo
  {volume} {33}},\ \bibinfo {pages} {135009} (\bibinfo {year} {2016})},\
  \Eprint {https://arxiv.org/abs/1404.7036} {arXiv:1404.7036 [gr-qc]}
  \BibitemShut {NoStop}%
\bibitem [{\citenamefont {Cardoso}\ \emph {et~al.}(2014)\citenamefont
  {Cardoso}, \citenamefont {Crispino}, \citenamefont {Macedo}, \citenamefont
  {Okawa},\ and\ \citenamefont {Pani}}]{CardosoEtAl2014}%
  \BibitemOpen
  \bibfield  {author} {\bibinfo {author} {\bibfnamefont {V.}~\bibnamefont
  {Cardoso}}, \bibinfo {author} {\bibfnamefont {L.~C.~B.}\ \bibnamefont
  {Crispino}}, \bibinfo {author} {\bibfnamefont {C.~F.~B.}\ \bibnamefont
  {Macedo}}, \bibinfo {author} {\bibfnamefont {H.}~\bibnamefont {Okawa}},\ and\
  \bibinfo {author} {\bibfnamefont {P.}~\bibnamefont {Pani}},\ }\href
  {https://doi.org/10.1103/PhysRevD.90.044069} {\bibfield  {journal} {\bibinfo
  {journal} {Phys. Rev. D}\ }\textbf {\bibinfo {volume} {90}},\ \bibinfo
  {pages} {044069} (\bibinfo {year} {2014})},\ \Eprint
  {https://arxiv.org/abs/1406.5510} {arXiv:1406.5510 [gr-qc]} \BibitemShut
  {NoStop}%
\bibitem [{\citenamefont {Cunha}\ \emph {et~al.}(2023)\citenamefont {Cunha},
  \citenamefont {Herdeiro}, \citenamefont {Radu},\ and\ \citenamefont
  {Sanchis-Gual}}]{CunhaEtAl2023}%
  \BibitemOpen
  \bibfield  {author} {\bibinfo {author} {\bibfnamefont {P.~V.~P.}\
  \bibnamefont {Cunha}}, \bibinfo {author} {\bibfnamefont {C.}~\bibnamefont
  {Herdeiro}}, \bibinfo {author} {\bibfnamefont {E.}~\bibnamefont {Radu}},\
  and\ \bibinfo {author} {\bibfnamefont {N.}~\bibnamefont {Sanchis-Gual}},\
  }\href {https://doi.org/10.1103/PhysRevLett.130.061401} {\bibfield  {journal}
  {\bibinfo  {journal} {Phys. Rev. Lett.}\ }\textbf {\bibinfo {volume} {130}},\
  \bibinfo {pages} {061401} (\bibinfo {year} {2023})},\ \Eprint
  {https://arxiv.org/abs/2207.13713} {arXiv:2207.13713 [gr-qc]} \BibitemShut
  {NoStop}%
\bibitem [{\citenamefont {Cunha}(2025)}]{Cunha2025}%
  \BibitemOpen
  \bibfield  {author} {\bibinfo {author} {\bibfnamefont {P.~V.~P.}\
  \bibnamefont {Cunha}},\ }\href
  {https://doi.org/10.1088/1475-7516/2025/05/083} {\bibfield  {journal}
  {\bibinfo  {journal} {JCAP}\ }\textbf {\bibinfo {volume} {05}},\ \bibinfo
  {pages} {083}},\ \Eprint {https://arxiv.org/abs/2503.00117} {arXiv:2503.00117
  [gr-qc]} \BibitemShut {NoStop}%
\bibitem [{\citenamefont {Cunha}\ \emph
  {et~al.}(2017{\natexlab{b}})\citenamefont {Cunha}, \citenamefont {Berti},\
  and\ \citenamefont {Herdeiro}}]{CunhaBertiHerdeiro2017}%
  \BibitemOpen
  \bibfield  {author} {\bibinfo {author} {\bibfnamefont {P.~V.~P.}\
  \bibnamefont {Cunha}}, \bibinfo {author} {\bibfnamefont {E.}~\bibnamefont
  {Berti}},\ and\ \bibinfo {author} {\bibfnamefont {C.~A.~R.}\ \bibnamefont
  {Herdeiro}},\ }\href {https://doi.org/10.1103/PhysRevLett.119.251102}
  {\bibfield  {journal} {\bibinfo  {journal} {{Phys. Rev. Lett.}}\ }\textbf
  {\bibinfo {volume} {119}},\ \bibinfo {pages} {251102} (\bibinfo {year}
  {2017}{\natexlab{b}})},\ \Eprint {https://arxiv.org/abs/1708.04211}
  {arXiv:1708.04211 [gr-qc]} \BibitemShut {NoStop}%
\bibitem [{\citenamefont {Cunha}\ and\ \citenamefont
  {Herdeiro}(2020)}]{CunhaHerdeiro2020}%
  \BibitemOpen
  \bibfield  {author} {\bibinfo {author} {\bibfnamefont {P.~V.~P.}\
  \bibnamefont {Cunha}}\ and\ \bibinfo {author} {\bibfnamefont {C.~A.~R.}\
  \bibnamefont {Herdeiro}},\ }\href
  {https://doi.org/10.1103/PhysRevLett.124.181101} {\bibfield  {journal}
  {\bibinfo  {journal} {Phys. Rev. Lett.}\ }\textbf {\bibinfo {volume} {124}},\
  \bibinfo {pages} {181101} (\bibinfo {year} {2020})},\ \Eprint
  {https://arxiv.org/abs/2003.06445} {arXiv:2003.06445 [gr-qc]} \BibitemShut
  {NoStop}%
\bibitem [{\citenamefont {Xavier}\ \emph {et~al.}(2024)\citenamefont {Xavier},
  \citenamefont {Herdeiro},\ and\ \citenamefont
  {Crispino}}]{XavierHerdeiroCrispino2024}%
  \BibitemOpen
  \bibfield  {author} {\bibinfo {author} {\bibfnamefont {S.~V. M. C.~B.}\
  \bibnamefont {Xavier}}, \bibinfo {author} {\bibfnamefont {C.~A.~R.}\
  \bibnamefont {Herdeiro}},\ and\ \bibinfo {author} {\bibfnamefont {L.~C.~B.}\
  \bibnamefont {Crispino}},\ }\href
  {https://doi.org/10.1103/PhysRevD.109.124065} {\bibfield  {journal} {\bibinfo
   {journal} {Phys. Rev. D}\ }\textbf {\bibinfo {volume} {109}},\ \bibinfo
  {pages} {124065} (\bibinfo {year} {2024})},\ \Eprint
  {https://arxiv.org/abs/2404.02208} {arXiv:2404.02208 [gr-qc]} \BibitemShut
  {NoStop}%
\bibitem [{\citenamefont {Wei}(2020)}]{Wei2020}%
  \BibitemOpen
  \bibfield  {author} {\bibinfo {author} {\bibfnamefont {S.-W.}\ \bibnamefont
  {Wei}},\ }\href {https://doi.org/10.1103/PhysRevD.102.064039} {\bibfield
  {journal} {\bibinfo  {journal} {Phys. Rev. D}\ }\textbf {\bibinfo {volume}
  {102}},\ \bibinfo {pages} {064039} (\bibinfo {year} {2020})},\ \Eprint
  {https://arxiv.org/abs/2006.02112} {arXiv:2006.02112 [gr-qc]} \BibitemShut
  {NoStop}%
\bibitem [{\citenamefont {Ghosh}\ and\ \citenamefont
  {Sarkar}(2021)}]{GhoshSarkar2021}%
  \BibitemOpen
  \bibfield  {author} {\bibinfo {author} {\bibfnamefont {R.}~\bibnamefont
  {Ghosh}}\ and\ \bibinfo {author} {\bibfnamefont {S.}~\bibnamefont {Sarkar}},\
  }\href {https://doi.org/10.1103/PhysRevD.104.044019} {\bibfield  {journal}
  {\bibinfo  {journal} {Phys. Rev. D}\ }\textbf {\bibinfo {volume} {104}},\
  \bibinfo {pages} {044019} (\bibinfo {year} {2021})},\ \Eprint
  {https://arxiv.org/abs/2107.07370} {arXiv:2107.07370 [gr-qc]} \BibitemShut
  {NoStop}%
\bibitem [{\citenamefont {Cunha}\ \emph {et~al.}(2024)\citenamefont {Cunha},
  \citenamefont {Herdeiro},\ and\ \citenamefont
  {Novo}}]{CunhaHerdeiroNovo2024}%
  \BibitemOpen
  \bibfield  {author} {\bibinfo {author} {\bibfnamefont {P.~V.~P.}\
  \bibnamefont {Cunha}}, \bibinfo {author} {\bibfnamefont {C.~A.~R.}\
  \bibnamefont {Herdeiro}},\ and\ \bibinfo {author} {\bibfnamefont {J.~P.~A.}\
  \bibnamefont {Novo}},\ }\href {https://doi.org/10.1103/PhysRevD.109.064050}
  {\bibfield  {journal} {\bibinfo  {journal} {Phys. Rev. D}\ }\textbf {\bibinfo
  {volume} {109}},\ \bibinfo {pages} {064050} (\bibinfo {year} {2024})},\
  \Eprint {https://arxiv.org/abs/2401.05495} {arXiv:2401.05495 [gr-qc]}
  \BibitemShut {NoStop}%
\bibitem [{\citenamefont {Guo}\ and\ \citenamefont {Gao}(2021)}]{GuoGao2021}%
  \BibitemOpen
  \bibfield  {author} {\bibinfo {author} {\bibfnamefont {M.}~\bibnamefont
  {Guo}}\ and\ \bibinfo {author} {\bibfnamefont {S.}~\bibnamefont {Gao}},\
  }\href {https://doi.org/10.1103/PhysRevD.103.104031} {\bibfield  {journal}
  {\bibinfo  {journal} {Phys. Rev. D}\ }\textbf {\bibinfo {volume} {103}},\
  \bibinfo {pages} {104031} (\bibinfo {year} {2021})},\ \Eprint
  {https://arxiv.org/abs/2011.02211} {arXiv:2011.02211 [gr-qc]} \BibitemShut
  {NoStop}%
\bibitem [{\citenamefont {Hod}(2018)}]{Hod2018}%
  \BibitemOpen
  \bibfield  {author} {\bibinfo {author} {\bibfnamefont {S.}~\bibnamefont
  {Hod}},\ }\href {https://doi.org/10.1016/j.physletb.2017.11.021} {\bibfield
  {journal} {\bibinfo  {journal} {Phys. Lett. B}\ }\textbf {\bibinfo {volume}
  {776}},\ \bibinfo {pages} {1} (\bibinfo {year} {2018})},\ \Eprint
  {https://arxiv.org/abs/1710.00836} {arXiv:1710.00836 [gr-qc]} \BibitemShut
  {NoStop}%
\bibitem [{\citenamefont {Dolan}\ and\ \citenamefont
  {Shipley}(2016)}]{DolanShipley2016}%
  \BibitemOpen
  \bibfield  {author} {\bibinfo {author} {\bibfnamefont {S.~R.}\ \bibnamefont
  {Dolan}}\ and\ \bibinfo {author} {\bibfnamefont {J.~O.}\ \bibnamefont
  {Shipley}},\ }\href {https://doi.org/10.1103/PhysRevD.94.044038} {\bibfield
  {journal} {\bibinfo  {journal} {{Phys. Rev. D}}\ }\textbf {\bibinfo {volume}
  {94}},\ \bibinfo {pages} {044038} (\bibinfo {year} {2016})},\ \Eprint
  {https://arxiv.org/abs/1605.07193} {arXiv:1605.07193 [gr-qc]} \BibitemShut
  {NoStop}%
\bibitem [{\citenamefont {Shipley}(2023)}]{Shipley2023}%
  \BibitemOpen
  \bibfield  {author} {\bibinfo {author} {\bibfnamefont {J.~O.}\ \bibnamefont
  {Shipley}},\ }\href {https://doi.org/10.1103/PhysRevD.108.084040} {\bibfield
  {journal} {\bibinfo  {journal} {Phys. Rev. D}\ }\textbf {\bibinfo {volume}
  {108}},\ \bibinfo {pages} {084040} (\bibinfo {year} {2023})},\ \Eprint
  {https://arxiv.org/abs/2305.14833} {arXiv:2305.14833 [gr-qc]} \BibitemShut
  {NoStop}%
\bibitem [{\citenamefont {Claudel}\ \emph {et~al.}(2001)\citenamefont
  {Claudel}, \citenamefont {Virbhadra},\ and\ \citenamefont
  {Ellis}}]{ClaudelVirbhadraEllis2001}%
  \BibitemOpen
  \bibfield  {author} {\bibinfo {author} {\bibfnamefont {C.-M.}\ \bibnamefont
  {Claudel}}, \bibinfo {author} {\bibfnamefont {K.~S.}\ \bibnamefont
  {Virbhadra}},\ and\ \bibinfo {author} {\bibfnamefont {G.~F.~R.}\ \bibnamefont
  {Ellis}},\ }\href {https://doi.org/10.1063/1.1308507} {\bibfield  {journal}
  {\bibinfo  {journal} {J. Math. Phys.}\ }\textbf {\bibinfo {volume} {42}},\
  \bibinfo {pages} {818} (\bibinfo {year} {2001})},\ \Eprint
  {https://arxiv.org/abs/gr-qc/0005050} {arXiv:gr-qc/0005050} \BibitemShut
  {NoStop}%
\bibitem [{\citenamefont {Perlick}(2005)}]{Perlick2005}%
  \BibitemOpen
  \bibfield  {author} {\bibinfo {author} {\bibfnamefont {V.}~\bibnamefont
  {Perlick}},\ }\bibfield  {booktitle} {\emph {\bibinfo {booktitle} {{4th World
  Congress of Nonlinear Analysts (WCNA 2004)}}},\ }\href@noop {} {\  (\bibinfo
  {year} {2005})},\ \Eprint {https://arxiv.org/abs/gr-qc/0512066}
  {arXiv:gr-qc/0512066} \BibitemShut {NoStop}%
\bibitem [{\citenamefont {Koga}\ and\ \citenamefont
  {Harada}(2019)}]{KogaHarada2019}%
  \BibitemOpen
  \bibfield  {author} {\bibinfo {author} {\bibfnamefont {Y.}~\bibnamefont
  {Koga}}\ and\ \bibinfo {author} {\bibfnamefont {T.}~\bibnamefont {Harada}},\
  }\href {https://doi.org/10.1103/PhysRevD.100.064040} {\bibfield  {journal}
  {\bibinfo  {journal} {Phys. Rev. D}\ }\textbf {\bibinfo {volume} {100}},\
  \bibinfo {pages} {064040} (\bibinfo {year} {2019})},\ \Eprint
  {https://arxiv.org/abs/1907.07336} {arXiv:1907.07336 [gr-qc]} \BibitemShut
  {NoStop}%
\bibitem [{\citenamefont {Teo}(2003)}]{Teo2003}%
  \BibitemOpen
  \bibfield  {author} {\bibinfo {author} {\bibfnamefont {E.}~\bibnamefont
  {Teo}},\ }\href {https://doi.org/10.1023/A:1026286607562} {\bibfield
  {journal} {\bibinfo  {journal} {{Gen. Relativ. Gravitation}}\ }\textbf
  {\bibinfo {volume} {35}},\ \bibinfo {pages} {1909} (\bibinfo {year}
  {2003})}\BibitemShut {NoStop}%
\bibitem [{\citenamefont {Kobialko}\ and\ \citenamefont
  {Gal'tsov}(2020)}]{KobialkoGaltsov2020}%
  \BibitemOpen
  \bibfield  {author} {\bibinfo {author} {\bibfnamefont {K.~V.}\ \bibnamefont
  {Kobialko}}\ and\ \bibinfo {author} {\bibfnamefont {D.~V.}\ \bibnamefont
  {Gal'tsov}},\ }\href {https://doi.org/10.1140/epjc/s10052-020-8070-z}
  {\bibfield  {journal} {\bibinfo  {journal} {Eur. Phys. J. C}\ }\textbf
  {\bibinfo {volume} {80}},\ \bibinfo {pages} {527} (\bibinfo {year} {2020})},\
  \Eprint {https://arxiv.org/abs/2002.04280} {arXiv:2002.04280 [gr-qc]}
  \BibitemShut {NoStop}%
\bibitem [{\citenamefont {Papapetrou}(1966)}]{Papapetrou1966}%
  \BibitemOpen
  \bibfield  {author} {\bibinfo {author} {\bibfnamefont {A.}~\bibnamefont
  {Papapetrou}},\ }\href@noop {} {\bibfield  {journal} {\bibinfo  {journal}
  {Ann. Inst. H. Poincare Phys. Theor. A}\ }\textbf {\bibinfo {volume} {4}},\
  \bibinfo {pages} {83} (\bibinfo {year} {1966})}\BibitemShut {NoStop}%
\bibitem [{\citenamefont {Wald}(1984)}]{Wald1984}%
  \BibitemOpen
  \bibfield  {author} {\bibinfo {author} {\bibfnamefont {R.~M.}\ \bibnamefont
  {Wald}},\ }\href {https://doi.org/10.7208/chicago/9780226870373.001.0001}
  {\emph {\bibinfo {title} {{General Relativity}}}}\ (\bibinfo  {publisher}
  {University of Chicago Press},\ \bibinfo {year} {1984})\BibitemShut {NoStop}%
\bibitem [{\citenamefont {Babichev}\ and\ \citenamefont
  {Mazza}(2025)}]{BabichevMazza2025}%
  \BibitemOpen
  \bibfield  {author} {\bibinfo {author} {\bibfnamefont {E.}~\bibnamefont
  {Babichev}}\ and\ \bibinfo {author} {\bibfnamefont {J.}~\bibnamefont
  {Mazza}},\ }\href {https://doi.org/10.1088/1475-7516/2025/10/011} {\bibfield
  {journal} {\bibinfo  {journal} {JCAP}\ }\textbf {\bibinfo {volume} {10}},\
  \bibinfo {pages} {011}},\ \Eprint {https://arxiv.org/abs/2505.08880}
  {arXiv:2505.08880 [gr-qc]} \BibitemShut {NoStop}%
\bibitem [{\citenamefont {Kundt}\ and\ \citenamefont
  {Tr\"{u}mper}(1966)}]{KundtTrumper1966}%
  \BibitemOpen
  \bibfield  {author} {\bibinfo {author} {\bibfnamefont {W.}~\bibnamefont
  {Kundt}}\ and\ \bibinfo {author} {\bibfnamefont {M.}~\bibnamefont
  {Tr\"{u}mper}},\ }\href {https://doi.org/10.1007/BF01325677} {\bibfield
  {journal} {\bibinfo  {journal} {Z. Phys.}\ }\textbf {\bibinfo {volume}
  {192}},\ \bibinfo {pages} {419} (\bibinfo {year} {1966})}\BibitemShut
  {NoStop}%
\bibitem [{\citenamefont {Heusler}(1996)}]{Heusler1996}%
  \BibitemOpen
  \bibfield  {author} {\bibinfo {author} {\bibfnamefont {M.}~\bibnamefont
  {Heusler}},\ }\href {https://doi.org/10.1017/CBO9780511661396} {\emph
  {\bibinfo {title} {{Black Hole Uniqueness Theorems}}}}\ (\bibinfo
  {publisher} {Cambridge University Press},\ \bibinfo {year}
  {1996})\BibitemShut {NoStop}%
\bibitem [{\citenamefont {Chru\'{s}ciel}\ \emph {et~al.}(2012)\citenamefont
  {Chru\'{s}ciel}, \citenamefont {Lopes~Costa},\ and\ \citenamefont
  {Heusler}}]{ChruscielLopesCostaHeusler2012}%
  \BibitemOpen
  \bibfield  {author} {\bibinfo {author} {\bibfnamefont {P.~T.}\ \bibnamefont
  {Chru\'{s}ciel}}, \bibinfo {author} {\bibfnamefont {J.}~\bibnamefont
  {Lopes~Costa}},\ and\ \bibinfo {author} {\bibfnamefont {M.}~\bibnamefont
  {Heusler}},\ }\href {https://doi.org/10.12942/lrr-2012-7} {\bibfield
  {journal} {\bibinfo  {journal} {Living Rev. Rel.}\ }\textbf {\bibinfo
  {volume} {15}},\ \bibinfo {pages} {7} (\bibinfo {year} {2012})},\ \Eprint
  {https://arxiv.org/abs/1205.6112} {arXiv:1205.6112 [gr-qc]} \BibitemShut
  {NoStop}%
\bibitem [{\citenamefont {Gourgoulhon}\ and\ \citenamefont
  {Bonazzola}(1993)}]{GourgoulhonBonazzola1993}%
  \BibitemOpen
  \bibfield  {author} {\bibinfo {author} {\bibfnamefont {E.}~\bibnamefont
  {Gourgoulhon}}\ and\ \bibinfo {author} {\bibfnamefont {S.}~\bibnamefont
  {Bonazzola}},\ }\href {https://doi.org/10.1103/PhysRevD.48.2635} {\bibfield
  {journal} {\bibinfo  {journal} {Phys. Rev. D}\ }\textbf {\bibinfo {volume}
  {48}},\ \bibinfo {pages} {2635} (\bibinfo {year} {1993})}\BibitemShut
  {NoStop}%
\bibitem [{\citenamefont {Carter}(1969)}]{Carter1969}%
  \BibitemOpen
  \bibfield  {author} {\bibinfo {author} {\bibfnamefont {B.}~\bibnamefont
  {Carter}},\ }\href {https://doi.org/10.1063/1.1664763} {\bibfield  {journal}
  {\bibinfo  {journal} {J. Math. Phys.}\ }\textbf {\bibinfo {volume} {10}},\
  \bibinfo {pages} {70} (\bibinfo {year} {1969})}\BibitemShut {NoStop}%
\bibitem [{\citenamefont {Shipley}(2019)}]{Shipley2019}%
  \BibitemOpen
  \bibfield  {author} {\bibinfo {author} {\bibfnamefont {J.~O.}\ \bibnamefont
  {Shipley}},\ }\emph {\bibinfo {title} {Strong-field gravitational lensing by
  black holes}},\ \href {https://etheses.whiterose.ac.uk/24823/} {Ph.D.
  thesis},\ \bibinfo  {school} {University of Sheffield} (\bibinfo {year}
  {2019}),\ \Eprint {https://arxiv.org/abs/1909.04691} {arXiv:1909.04691
  [gr-qc]} \BibitemShut {NoStop}%
\bibitem [{\citenamefont {Chandrasekhar}(1983)}]{Chandrasekhar1983}%
  \BibitemOpen
  \bibfield  {author} {\bibinfo {author} {\bibfnamefont {S.}~\bibnamefont
  {Chandrasekhar}},\ }\href
  {https://doi.org/10.1093/oso/9780198503705.001.0001} {\emph {\bibinfo {title}
  {{The Mathematical Theory of Black Holes}}}},\ \bibinfo {series}
  {International Series of Monographs on Physics}, Vol.~\bibinfo {volume} {69}\
  (\bibinfo  {publisher} {Oxford University Press},\ \bibinfo {year}
  {1983})\BibitemShut {NoStop}%
\bibitem [{\citenamefont {Shipley}\ and\ \citenamefont
  {Dolan}(2016)}]{ShipleyDolan2016}%
  \BibitemOpen
  \bibfield  {author} {\bibinfo {author} {\bibfnamefont {J.~O.}\ \bibnamefont
  {Shipley}}\ and\ \bibinfo {author} {\bibfnamefont {S.~R.}\ \bibnamefont
  {Dolan}},\ }\href {https://doi.org/10.1088/0264-9381/33/17/175001} {\bibfield
   {journal} {\bibinfo  {journal} {Class. Quant. Grav.}\ }\textbf {\bibinfo
  {volume} {33}},\ \bibinfo {pages} {175001} (\bibinfo {year} {2016})},\
  \Eprint {https://arxiv.org/abs/1603.04469} {arXiv:1603.04469 [gr-qc]}
  \BibitemShut {NoStop}%
\bibitem [{\citenamefont {Weyl}(1917)}]{Weyl1917}%
  \BibitemOpen
  \bibfield  {author} {\bibinfo {author} {\bibfnamefont {H.}~\bibnamefont
  {Weyl}},\ }\href {https://doi.org/10.1002/andp.19173591804} {\bibfield
  {journal} {\bibinfo  {journal} {{Ann. Phys.}}\ }\textbf {\bibinfo {volume}
  {359}},\ \bibinfo {pages} {117} (\bibinfo {year} {1917})}\BibitemShut
  {NoStop}%
\bibitem [{\citenamefont {Griffiths}\ and\ \citenamefont
  {Podolsk\'{y}}(2009)}]{GriffithsPodolsky2009}%
  \BibitemOpen
  \bibfield  {author} {\bibinfo {author} {\bibfnamefont {J.~B.}\ \bibnamefont
  {Griffiths}}\ and\ \bibinfo {author} {\bibfnamefont {J.}~\bibnamefont
  {Podolsk\'{y}}},\ }\href {https://doi.org/10.1017/CBO9780511635397} {\emph
  {\bibinfo {title} {{Exact Space-Times in Einstein's General Relativity}}}},\
  Cambridge Monographs on Mathematical Physics\ (\bibinfo  {publisher}
  {Cambridge University Press},\ \bibinfo {address} {Cambridge},\ \bibinfo
  {year} {2009})\BibitemShut {NoStop}%
\bibitem [{\citenamefont {Yoshida}\ and\ \citenamefont
  {Eriguchi}(1997)}]{YoshidaEriguchi1997}%
  \BibitemOpen
  \bibfield  {author} {\bibinfo {author} {\bibfnamefont {S.}~\bibnamefont
  {Yoshida}}\ and\ \bibinfo {author} {\bibfnamefont {Y.}~\bibnamefont
  {Eriguchi}},\ }\href {https://doi.org/10.1103/PhysRevD.56.762} {\bibfield
  {journal} {\bibinfo  {journal} {Phys. Rev. D}\ }\textbf {\bibinfo {volume}
  {56}},\ \bibinfo {pages} {762} (\bibinfo {year} {1997})}\BibitemShut
  {NoStop}%
\bibitem [{\citenamefont {Kleihaus}\ \emph {et~al.}(2005)\citenamefont
  {Kleihaus}, \citenamefont {Kunz},\ and\ \citenamefont
  {List}}]{KleihausKunzList2005}%
  \BibitemOpen
  \bibfield  {author} {\bibinfo {author} {\bibfnamefont {B.}~\bibnamefont
  {Kleihaus}}, \bibinfo {author} {\bibfnamefont {J.}~\bibnamefont {Kunz}},\
  and\ \bibinfo {author} {\bibfnamefont {M.}~\bibnamefont {List}},\ }\href
  {https://doi.org/10.1103/PhysRevD.72.064002} {\bibfield  {journal} {\bibinfo
  {journal} {Phys. Rev. D}\ }\textbf {\bibinfo {volume} {72}},\ \bibinfo
  {pages} {064002} (\bibinfo {year} {2005})},\ \Eprint
  {https://arxiv.org/abs/gr-qc/0505143} {arXiv:gr-qc/0505143} \BibitemShut
  {NoStop}%
\bibitem [{\citenamefont {Grandcl{\'e}ment}(2017)}]{Grandclement2017}%
  \BibitemOpen
  \bibfield  {author} {\bibinfo {author} {\bibfnamefont {P.}~\bibnamefont
  {Grandcl{\'e}ment}},\ }\href {https://doi.org/10.1103/PhysRevD.95.084011}
  {\bibfield  {journal} {\bibinfo  {journal} {Phys. Rev. D}\ }\textbf {\bibinfo
  {volume} {95}},\ \bibinfo {pages} {084011} (\bibinfo {year} {2017})},\
  \Eprint {https://arxiv.org/abs/1612.07507} {arXiv:1612.07507 [gr-qc]}
  \BibitemShut {NoStop}%
\bibitem [{\citenamefont {Kaup}(1968)}]{Kaup1968}%
  \BibitemOpen
  \bibfield  {author} {\bibinfo {author} {\bibfnamefont {D.~J.}\ \bibnamefont
  {Kaup}},\ }\href {https://doi.org/10.1103/PhysRev.172.1331} {\bibfield
  {journal} {\bibinfo  {journal} {Phys. Rev.}\ }\textbf {\bibinfo {volume}
  {172}},\ \bibinfo {pages} {1331} (\bibinfo {year} {1968})}\BibitemShut
  {NoStop}%
\bibitem [{\citenamefont {Ruffini}\ and\ \citenamefont
  {Bonazzola}(1969)}]{RuffiniBonazzola1969}%
  \BibitemOpen
  \bibfield  {author} {\bibinfo {author} {\bibfnamefont {R.}~\bibnamefont
  {Ruffini}}\ and\ \bibinfo {author} {\bibfnamefont {S.}~\bibnamefont
  {Bonazzola}},\ }\href {https://doi.org/10.1103/PhysRev.187.1767} {\bibfield
  {journal} {\bibinfo  {journal} {Phys. Rev.}\ }\textbf {\bibinfo {volume}
  {187}},\ \bibinfo {pages} {1767} (\bibinfo {year} {1969})}\BibitemShut
  {NoStop}%
\bibitem [{\citenamefont {Schunck}\ and\ \citenamefont
  {Mielke}(2003)}]{SchunckMielke2003}%
  \BibitemOpen
  \bibfield  {author} {\bibinfo {author} {\bibfnamefont {F.~E.}\ \bibnamefont
  {Schunck}}\ and\ \bibinfo {author} {\bibfnamefont {E.~W.}\ \bibnamefont
  {Mielke}},\ }\href {https://doi.org/10.1088/0264-9381/20/20/201} {\bibfield
  {journal} {\bibinfo  {journal} {Class. Quant. Grav.}\ }\textbf {\bibinfo
  {volume} {20}},\ \bibinfo {pages} {R301} (\bibinfo {year} {2003})},\ \Eprint
  {https://arxiv.org/abs/0801.0307} {arXiv:0801.0307 [astro-ph]} \BibitemShut
  {NoStop}%
\bibitem [{\citenamefont {Herdeiro}\ and\ \citenamefont
  {Radu}(2014)}]{HerdeiroRadu2014}%
  \BibitemOpen
  \bibfield  {author} {\bibinfo {author} {\bibfnamefont {C.~A.~R.}\
  \bibnamefont {Herdeiro}}\ and\ \bibinfo {author} {\bibfnamefont
  {E.}~\bibnamefont {Radu}},\ }\href
  {https://doi.org/10.1103/PhysRevLett.112.221101} {\bibfield  {journal}
  {\bibinfo  {journal} {Phys. Rev. Lett.}\ }\textbf {\bibinfo {volume} {112}},\
  \bibinfo {pages} {221101} (\bibinfo {year} {2014})},\ \Eprint
  {https://arxiv.org/abs/1403.2757} {arXiv:1403.2757 [gr-qc]} \BibitemShut
  {NoStop}%
\bibitem [{\citenamefont {Cunha}\ \emph {et~al.}(2015)\citenamefont {Cunha},
  \citenamefont {Herdeiro}, \citenamefont {Radu},\ and\ \citenamefont
  {Runarsson}}]{CunhaEtAl2015}%
  \BibitemOpen
  \bibfield  {author} {\bibinfo {author} {\bibfnamefont {P.~V.~P.}\
  \bibnamefont {Cunha}}, \bibinfo {author} {\bibfnamefont {C.~A.~R.}\
  \bibnamefont {Herdeiro}}, \bibinfo {author} {\bibfnamefont {E.}~\bibnamefont
  {Radu}},\ and\ \bibinfo {author} {\bibfnamefont {H.~F.}\ \bibnamefont
  {Runarsson}},\ }\href {https://doi.org/10.1103/PhysRevLett.115.211102}
  {\bibfield  {journal} {\bibinfo  {journal} {Phys. Rev. Lett.}\ }\textbf
  {\bibinfo {volume} {115}},\ \bibinfo {pages} {211102} (\bibinfo {year}
  {2015})},\ \Eprint {https://arxiv.org/abs/1509.00021} {arXiv:1509.00021
  [gr-qc]} \BibitemShut {NoStop}%
\bibitem [{\citenamefont {Brito}\ \emph {et~al.}(2016)\citenamefont {Brito},
  \citenamefont {Cardoso}, \citenamefont {Herdeiro},\ and\ \citenamefont
  {Radu}}]{BritoEtAl2016}%
  \BibitemOpen
  \bibfield  {author} {\bibinfo {author} {\bibfnamefont {R.}~\bibnamefont
  {Brito}}, \bibinfo {author} {\bibfnamefont {V.}~\bibnamefont {Cardoso}},
  \bibinfo {author} {\bibfnamefont {C.~A.~R.}\ \bibnamefont {Herdeiro}},\ and\
  \bibinfo {author} {\bibfnamefont {E.}~\bibnamefont {Radu}},\ }\href
  {https://doi.org/10.1016/j.physletb.2015.11.051} {\bibfield  {journal}
  {\bibinfo  {journal} {Phys. Lett. B}\ }\textbf {\bibinfo {volume} {752}},\
  \bibinfo {pages} {291} (\bibinfo {year} {2016})},\ \Eprint
  {https://arxiv.org/abs/1508.05395} {arXiv:1508.05395 [gr-qc]} \BibitemShut
  {NoStop}%
\bibitem [{\citenamefont {Herdeiro}\ \emph {et~al.}(2016)\citenamefont
  {Herdeiro}, \citenamefont {Radu},\ and\ \citenamefont
  {R{\'u}narsson}}]{HerdeiroRaduRunarsson2016}%
  \BibitemOpen
  \bibfield  {author} {\bibinfo {author} {\bibfnamefont {C.}~\bibnamefont
  {Herdeiro}}, \bibinfo {author} {\bibfnamefont {E.}~\bibnamefont {Radu}},\
  and\ \bibinfo {author} {\bibfnamefont {H.}~\bibnamefont {R{\'u}narsson}},\
  }\href {https://doi.org/10.1088/0264-9381/33/15/154001} {\bibfield  {journal}
  {\bibinfo  {journal} {Class. Quant. Grav.}\ }\textbf {\bibinfo {volume}
  {33}},\ \bibinfo {pages} {154001} (\bibinfo {year} {2016})},\ \Eprint
  {https://arxiv.org/abs/1603.02687} {arXiv:1603.02687 [gr-qc]} \BibitemShut
  {NoStop}%
\bibitem [{\citenamefont {Sengo}\ \emph {et~al.}(2023)\citenamefont {Sengo},
  \citenamefont {Cunha}, \citenamefont {Herdeiro},\ and\ \citenamefont
  {Radu}}]{SengoEtAl2023}%
  \BibitemOpen
  \bibfield  {author} {\bibinfo {author} {\bibfnamefont {I.}~\bibnamefont
  {Sengo}}, \bibinfo {author} {\bibfnamefont {P.~V.~P.}\ \bibnamefont {Cunha}},
  \bibinfo {author} {\bibfnamefont {C.~A.~R.}\ \bibnamefont {Herdeiro}},\ and\
  \bibinfo {author} {\bibfnamefont {E.}~\bibnamefont {Radu}},\ }\href
  {https://doi.org/10.1088/1475-7516/2023/01/047} {\bibfield  {journal}
  {\bibinfo  {journal} {JCAP}\ }\textbf {\bibinfo {volume} {2023}}\bibfield
  {number} {\bibinfo  {number} { (01)},\ \bibinfo {pages} {047}},\ }\Eprint
  {https://arxiv.org/abs/2209.06237} {arXiv:2209.06237 [gr-qc]} \BibitemShut
  {NoStop}%
\bibitem [{\citenamefont {Sachs}(1962)}]{Sachs1962}%
  \BibitemOpen
  \bibfield  {author} {\bibinfo {author} {\bibfnamefont {R.~K.}\ \bibnamefont
  {Sachs}},\ }\href {https://doi.org/10.1098/rspa.1962.0206} {\bibfield
  {journal} {\bibinfo  {journal} {Proc. Roy. Soc. Lond. A}\ }\textbf {\bibinfo
  {volume} {270}},\ \bibinfo {pages} {103} (\bibinfo {year}
  {1962})}\BibitemShut {NoStop}%
\bibitem [{\citenamefont {Qiao}\ \emph {et~al.}(2025)\citenamefont {Qiao},
  \citenamefont {Li}, \citenamefont {Xie},\ and\ \citenamefont
  {Guo}}]{Qiao2025}%
  \BibitemOpen
  \bibfield  {author} {\bibinfo {author} {\bibfnamefont {C.}~\bibnamefont
  {Qiao}}, \bibinfo {author} {\bibfnamefont {M.}~\bibnamefont {Li}}, \bibinfo
  {author} {\bibfnamefont {D.}~\bibnamefont {Xie}},\ and\ \bibinfo {author}
  {\bibfnamefont {M.}~\bibnamefont {Guo}},\ }\href@noop {} {\  (\bibinfo {year}
  {2025})},\ \Eprint {https://arxiv.org/abs/2512.20802} {arXiv:2512.20802
  [gr-qc]} \BibitemShut {NoStop}%
\bibitem [{\citenamefont {Igata}\ and\ \citenamefont
  {Tomizawa}(2020)}]{IgataTomizawa2020}%
  \BibitemOpen
  \bibfield  {author} {\bibinfo {author} {\bibfnamefont {T.}~\bibnamefont
  {Igata}}\ and\ \bibinfo {author} {\bibfnamefont {S.}~\bibnamefont
  {Tomizawa}},\ }\href {https://doi.org/10.1103/PhysRevD.102.084003} {\bibfield
   {journal} {\bibinfo  {journal} {Phys. Rev. D}\ }\textbf {\bibinfo {volume}
  {102}},\ \bibinfo {pages} {084003} (\bibinfo {year} {2020})},\ \Eprint
  {https://arxiv.org/abs/2008.00179} {arXiv:2008.00179 [hep-th]} \BibitemShut
  {NoStop}%
\bibitem [{\citenamefont {Tavlayan}\ and\ \citenamefont
  {Tekin}(2023)}]{TavlayanTekin2023}%
  \BibitemOpen
  \bibfield  {author} {\bibinfo {author} {\bibfnamefont {A.}~\bibnamefont
  {Tavlayan}}\ and\ \bibinfo {author} {\bibfnamefont {B.}~\bibnamefont
  {Tekin}},\ }\href {https://doi.org/10.1103/PhysRevD.107.024016} {\bibfield
  {journal} {\bibinfo  {journal} {Phys. Rev. D}\ }\textbf {\bibinfo {volume}
  {107}},\ \bibinfo {pages} {024016} (\bibinfo {year} {2023})},\ \Eprint
  {https://arxiv.org/abs/2209.14873} {arXiv:2209.14873 [gr-qc]} \BibitemShut
  {NoStop}%
\bibitem [{\citenamefont {Cunha}\ \emph
  {et~al.}(2017{\natexlab{c}})\citenamefont {Cunha}, \citenamefont {Herdeiro},\
  and\ \citenamefont {Radu}}]{CunhaHerdeiroRadu2017}%
  \BibitemOpen
  \bibfield  {author} {\bibinfo {author} {\bibfnamefont {P.~V.~P.}\
  \bibnamefont {Cunha}}, \bibinfo {author} {\bibfnamefont {C.~A.~R.}\
  \bibnamefont {Herdeiro}},\ and\ \bibinfo {author} {\bibfnamefont
  {E.}~\bibnamefont {Radu}},\ }\href
  {https://doi.org/10.1103/PhysRevD.96.024039} {\bibfield  {journal} {\bibinfo
  {journal} {{Phys. Rev. D}}\ }\textbf {\bibinfo {volume} {96}},\ \bibinfo
  {pages} {024039} (\bibinfo {year} {2017}{\natexlab{c}})},\ \Eprint
  {https://arxiv.org/abs/1705.05461} {arXiv:1705.05461 [gr-qc]} \BibitemShut
  {NoStop}%
\bibitem [{\citenamefont {Lima~Junior}\ \emph {et~al.}(2021)\citenamefont
  {Lima~Junior}, \citenamefont {Cunha}, \citenamefont {Herdeiro},\ and\
  \citenamefont {Crispino}}]{LimaJuniorEtAl2021}%
  \BibitemOpen
  \bibfield  {author} {\bibinfo {author} {\bibfnamefont {H.~C.~D.}\
  \bibnamefont {Lima~Junior}}, \bibinfo {author} {\bibfnamefont {P.~V.~P.}\
  \bibnamefont {Cunha}}, \bibinfo {author} {\bibfnamefont {C.~A.~R.}\
  \bibnamefont {Herdeiro}},\ and\ \bibinfo {author} {\bibfnamefont {L.~C.~B.}\
  \bibnamefont {Crispino}},\ }\href
  {https://doi.org/10.1103/PhysRevD.104.044018} {\bibfield  {journal} {\bibinfo
   {journal} {{Phys. Rev. D}}\ }\textbf {\bibinfo {volume} {104}},\ \bibinfo
  {pages} {044018} (\bibinfo {year} {2021})},\ \Eprint
  {https://arxiv.org/abs/2104.09577} {arXiv:2104.09577 [gr-qc]} \BibitemShut
  {NoStop}%
\end{thebibliography}%

\end{document}